\documentclass[twocolumn]{aastex62}
\usepackage{color,soul}
\usepackage{comment}
\usepackage{amsmath}
\usepackage{tablefootnote}

\newcommand{\nm}{\mathbin{\phantom{-}}}

\defcitealias{Bell2003}{B03}
\defcitealias{Into2013}{IP13}
\defcitealias{Kouroumpatzakis2023}{K23}
\defcitealias{Jarrett2023}{J23}

\graphicspath{{./}{figures/}}

\submitjournal{ApJ}

\begin{document}

\title{Stellar Mass Calibrations for Local Low-Mass Galaxies}

\correspondingauthor{Mithi A. C. de los Reyes}
\email{mdelosreyes@amherst.edu}

\author[0000-0002-4739-046X]{Mithi A. C. de los Reyes}
\affiliation{Department of Physics \& Astronomy, Amherst College, 6 East Drive, Amherst, MA 01002, USA}

\author[0000-0002-8320-2198]{Yasmeen Asali}
\affiliation{Department of Astronomy, Yale University, New Haven, CT 06520, USA}

\author[0000-0003-2229-011X]{Risa H. Wechsler}
\affiliation{Kavli Institute for Particle Astrophysics \& Cosmology, Stanford University, 452 Lomita Mall, Stanford, CA 94305, USA}
\affiliation{Department of Physics, Stanford University, 382 Via Pueblo Mall, Stanford, CA 94305, USA}
\affiliation{SLAC National Accelerator Laboratory, 2575 Sand Hill Road, Menlo Park, CA 94025, USA}

\author{Marla Geha}
\affiliation{Department of Astronomy, Yale University, New Haven, CT 06520, USA}

\author[0000-0002-1200-0820]{Yao-Yuan Mao}
\affiliation{Department of Physics and Astronomy, University of Utah, Salt Lake City, UT 84112, USA}

\author{Erin Kado-Fong}
\affiliation{Department of Astronomy, Yale University, New Haven, CT 06520, USA}

\author[0000-0002-4940-3009]{Ragadeepika Pucha}
\affiliation{Department of Physics and Astronomy, University of Utah, Salt Lake City, UT 84112, USA}

\author{William Grant}
\affiliation{Department of Physics \& Astronomy, Amherst College, 6 East Drive, Amherst, MA 01002, USA}

\author[0000-0003-0965-605X]{Pratik J. Gandhi}
\affiliation{Department of Astronomy, Yale University, New Haven, CT 06520, USA}
\affiliation{Department of Physics and Astronomy, University of California - Davis, One Shields Avenue, Davis, CA 95616, USA}

\author{Viraj Manwadkar}
\affiliation{Kavli Institute for Particle Astrophysics \& Cosmology, Stanford University, 452 Lomita Mall,  Stanford, CA 94305, USA}
\affiliation{Department of Physics, Stanford University, 382 Via Pueblo Mall, Stanford, CA 94305, USA}
\affiliation{SLAC National Accelerator Laboratory, 2575 Sand Hill Road, Menlo Park, CA 94025, USA}

\author{Anna Engelhardt}
\affiliation{Department of Physics, George Mason University, 4400 University Dr, Fairfax, VA 22030, USA}

\author{Ferah Munshi}
\affiliation{Department of Physics, George Mason University, 4400 University Dr, Fairfax, VA 22030, USA}

\author{Yunchong Wang}
\affiliation{Kavli Institute for Particle Astrophysics \& Cosmology, Stanford University, 452 Lomita Mall,  Stanford, CA 94305, USA}
\affiliation{Department of Physics, Stanford University, 382 Via Pueblo Mall, Stanford, CA 94305, USA}
\affiliation{SLAC National Accelerator Laboratory, 2575 Sand Hill Road, Menlo Park, CA 94025, USA}

\begin{abstract}
The stellar masses of galaxies are measured from integrated light via several methods --- however, few of these methods were designed for low-mass ($M_{\star}\lesssim10^{8}~\rm{M_{\odot}}$) ``dwarf'' galaxies, whose properties (e.g., stochastic star formation, low metallicity) pose unique challenges for estimating stellar masses.
In this work, we quantify the precision and accuracy at which stellar masses of low-mass galaxies can be recovered using UV/optical/IR photometry.
We use mock observations of 469 low-mass galaxies from a variety of models, including both semi-empirical models (GRUMPY and UniverseMachine-SAGA) and cosmological baryonic zoom-in simulations (MARVELous Dwarfs and FIRE-2), to test literature color--$M_\star/L$ relations and multi-wavelength spectral energy distribution (SED) mass estimators.
We identify a list of ``best practices'' for measuring stellar masses of low-mass galaxies from integrated photometry.
We find that literature color--$M_\star/L$ relations are often unable to capture the bursty star formation histories (SFHs) of low-mass galaxies, and we develop an updated prescription for stellar mass based on $g-r$ color that is better able to recover stellar masses for the bursty low-mass galaxies in our sample (with $\sim0.1$~dex precision).
SED fitting can also precisely recover stellar masses of low-mass galaxies, but this requires thoughtful choices about the form of the assumed SFH: parametric SFHs can underestimate stellar mass by as much as $\sim0.4$~dex, while non-parametric SFHs recover true stellar masses with insignificant offset ($-0.03\pm0.11~$dex).
Finally, we also caution that non-informative (wide) dust attenuation priors may introduce $M_\star$ uncertainties of up to $\sim0.6$~dex.
\end{abstract}

\section{Introduction} 
\label{sec:intro}

Much of observational astronomy can be distilled into a single exercise: converting the measured light from astrophysical objects to physical properties.
The mass of a galaxy is one of its most fundamental properties, and one that heavily influences not only a galaxy's observable properties, but its evolution over cosmic time.
The \emph{stellar} mass ($M_{\star}$) of a galaxy is a particularly useful property that encodes the galaxy's cumulative star formation, merger, and accretion history.
As a result, stellar mass is frequently used as the independent variable in empirical relationships, such as the stellar mass-halo mass relation \citep[e.g.,][]{Moster2010,Wechsler2018,Girelli2020}, the star formation main sequence \citep[e.g.,][]{Brinchmann2004,Popesso2023}, and the mass--metallicity relation \citep[e.g.,][]{Tremonti2004,Maiolino2019}, which are in turn used to probe galaxy evolution processes.

Measuring the stellar mass of a galaxy from its integrated light --- the only option for most galaxies, which are too distant to resolve individual stars --- is a non-trivial exercise. 
To the first order, a galaxy's stellar mass is correlated with its total luminosity: more stars produce more light.
However, determining the exact conversion from light to stellar mass (i.e., the mass-to-light ratio, $M_{\star}/L$) is complicated by several factors.

For a single coeval stellar population, the most massive stars contribute the majority of light, while the more numerous lower-mass stars contribute the majority of mass; the ratio of high- to low-mass stars is set at birth by the stellar initial mass function (IMF) and modified by stellar evolution.
A galaxy consists of multiple such stellar populations with varying ages and metallicities. 
The light from these populations is also modified by baryonic components like gas and dust located both within the galaxy and along the observed line of sight.
To estimate $M_{\star}/L$, one can model each of these components --- the stellar populations and their evolution, nebular emission, and dust attenuation and emission --- to obtain a model for a galaxy's observed spectrum.
This is the basis of stellar population synthesis (SPS) modeling \citep{Tinsley1968}, the most common method of estimating a galaxy's stellar mass.
SPS models are fit to the galaxy's observed light, which are typically in the form of (1) single fluxes or colors, (2) broadband spectral energy distributions (SEDs), or (3) spectra \citep[for a review, see][]{Conroy2013}.
Measurements of $M_\star$ therefore depend on all of the assumptions implicit in SPS modeling: the observable properties of a single stellar population with a given age and metallicity (which depends on the form of the IMF and stellar evolution models), the stellar populations that comprise a galaxy (which depends on the galaxy's assumed star formation history or SFH), and the impact of gas and dust (which depends on heavily simplified models).

Several works have tested the effect of these assumptions on $M_\star$ measurements; in particular, quantifying the uncertainties from SED fitting has become especially relevant in the era of JWST, which has begun to obtain SEDs of galaxies in the early universe. % \citep[e.g.,][]{XXX}.
For example, \citet{Lower2020} tested the impact of different SFH assumptions when measuring $M_\star$ from SED fitting, and \citet{Pacifici2023} tested the effect of using different SED fitting codes.
However, nearly all of these tests have focused on relatively massive galaxies: although \citet{Lower2020}'s sample included galaxies down to $10^{7.6}~\rm{M_\odot}$, the bulk of their sample was in the higher mass range of $10^{8-12}~\rm{M_\odot}$, and \citet{Pacifici2023} did not consider galaxies below $10^8~\rm{M_\odot}$ at all.

Low-mass ``dwarf'' galaxies ($M_\star \lesssim 10^8~\rm{M_\odot}$) have often been overlooked in these tests,
partly because they pose unique problems for SPS modeling.
For example, stellar populations in low-mass galaxies are strongly dependent on stochastic sampling of the IMF \citep[e.g.,][]{Applebaum2020}, as well as on low metallicities \citep[which can influence the evolution of binary and massive stars, e.g.,][]{Klencki2020}. 
Low metallicities may also affect the abundance and properties of dust \citep[e.g.,][]{Hamanowicz2024}, making it difficult to ``de-redden'' the observed light from low-mass galaxies. 
Finally, low-mass galaxies are thought to have relatively complex, stochastic star formation histories \citep{Emami2019}, making their SEDs and spectra more difficult to fit with SPS models.

However, low-mass galaxies are the most numerous type of galaxy in the universe \citep{Schechter1976}.
Due to their small gravitational potentials, they are sensitive to both internal processes like stellar feedback \citep{Collins2022} and external environmental effects \citep[e.g.,][]{Dekel2003,Farouki1981,Deason2014}, making them excellent testbeds for cosmological and galaxy evolution models.
With the advent of a new generation of extragalactic surveys --- e.g., the Dark Energy Spectroscopic Instrument \citep[DESI;][]{DESI} and DESI Legacy Surveys \citep{Dey2019}, the Legacy Survey of Space and Time \citep[LSST;][]{LSST2019}, Euclid \citep{Euclid}, and Roman \citep{Roman} --- we are beginning to observe unprecedentedly large samples of low-mass galaxies across a range of environments.
Accurately characterizing the masses of these galaxies is of critical importance: the low-mass end of the galaxy mass function can constrain cosmological models \citep{Sales2022}, and extending scaling relations to low masses can constrain galaxy physics \citep[e.g.,][]{Woo2008}.

In this work, we test a number of stellar mass measurement techniques on simulated low-mass galaxies ($M_\star=10^{4-8}~\rm{M_\odot}$).
In light of several ongoing and upcoming galaxy surveys, we primarily focus on methods that use UV/optical/IR photometry.
We identify some best practices for measuring low-mass galaxy masses, and we develop statistical corrections to extend commonly used empirical calibrations to the low-mass regime.

The structure of this paper is as follows.
We present the stellar mass calibrations used in this paper in Section~\ref{sec:calibrations}, and we describe the observed and simulated datasets used to test these stellar mass calibrations in Section~\ref{sec:models}. 
We then test the recovery of ``ground-truth'' stellar masses from simulated data: using photometric calibrations in Section~\ref{sec:results_photcalib}, and using SED fitting in Section~\ref{sec:results_sed}. 
We summarize our main results and list ``best practices'' for measuring low-mass galaxy masses in Section~\ref{sec:bestpractices}.
In Section~\ref{sec:discussion}, we discuss potential systematics in our experiment, and we summarize our conclusions in Section~\ref{sec:summary}.

\section{Stellar Mass Calibrations} 
\label{sec:calibrations}

Here we describe the stellar mass measurement techniques considered in this work. 
We emphasize that this is not an exhaustive list, particularly since we only focus on one SED fitting code (Section~\ref{sec:prospector}).
We convert all calibrations to a \cite{Chabrier2003} IMF.

\subsection{Empirical photometric calibrations}
\label{sec:fluxcalibs}

We first consider empirical calibrations based on optical and IR photometry.
Such approaches are attractive from an observational perspective because measuring a galaxy's flux in one or two photometric bands requires far fewer resources than measuring a multi-band SED over a broad wavelength range, let alone a spectrum.
Fortunately, several previous studies have found relatively tight correlations between stellar mass (typically measured from full SED fitting) and combinations of different photometric bands.
Historically, these have often been parameterized as relationships between $M_{\star}/L$ and a single color \citep[e.g.,][]{BelldeJong2001}, but a number of studies have developed stellar mass calibrations based on anywhere from one to three photometric bands.
In this work, we use the following calibrations.

\subsubsection{Optical $M_{\star}/L$ calibrations}
While there are many optical-NIR color-$M_\star/L$ relations \citep[e.g.,][]{Portinari2004,Zibetti2009,Gallazzi2009,Taylor2011,Roediger2015,Schombert2019,Kouroumpatzakis2023}, we have chosen to focus first on the color-$M_{\star}/L$ relations by \citet{Bell2003} and \citet{Into2013} because these have been used in a number of recent low-mass galaxy surveys.
For example, the Satellites Around Galactic Analogs survey \citep[SAGA;][]{Geha2017} and other surveys of low-metallicity/low-mass galaxies \citep{Hsyu2018} use recalibrations of the \citet{Bell2003} relations to measure stellar masses, while the Exploration of Local VolumE Satellites survey \citep[ELVES;][]{Carlsten2022,Danieli2023} use a color-$M_\star/L$ relation from \citet{Into2013}.

In the following equations in this section, all optical magnitudes are AB magnitudes that have been both corrected for Milky Way extinction and $k$-corrected to $z=0$ using $g-r$ color \citep{Chilingarian2010}.
The corrected $r$- and $g$-band magnitudes are denoted as $M_{r,0}$ and $M_{g,0}$.
Unless otherwise noted, we assume absolute solar magnitudes from \citet{Willmer2018}: 4.65 for $r$-band and 5.11 for $g$-band.

\cite{Bell2003} developed color-$M_{\star}/L$ relations based on optical-NIR SEDs of galaxies with photometry from the Two Micron All Sky Survey (2MASS) and the Sloan Digital Sky Survey (SDSS).
We use their $g-r$ relation: 
\begin{equation}
\label{eq:gr}
    \log(M_{\star}/\rm{M_{\odot}}) = 1.461 + 1.098(g-r)_0 - 0.4M_{r,0}.
\end{equation}
This relation was calibrated using a sample of galaxies with masses down to $\sim10^{8.5}~\rm{M_{\odot}}$.
To fit these galaxies, \cite{Bell2003} constructed a grid of SPS models with metallicities between $10^{-2.3}<Z/\rm{Z_\odot}<10^{0.4}$\footnote{This metallicity range is wide enough to include the expected metallicities of low-mass galaxies, based on the mass-metallicity relation (Figure~\ref{fig:mz_sim}).}.
They assumed exponentially decaying SFHs and did not account for dust attenuation, although they ran some simple tests to quantify the impact of adding star formation bursts and a simple constant dust reddening model.

\citet{Into2013} also proposed color-$M_\star/L$ relations using more detailed asymptotic giant branch models.
Here we consider the relation between $g-r$ and $M_\star/L_{g}$ used by the ELVES survey \citep{Danieli2023}:
\begin{equation}
\label{eq:gr_ip}
    \log(M_{\star}/\rm{M_{\odot}}) = 1.332 + 1.774(g-r)_0 - 0.4M_{g,0}
\end{equation}
Note that this equation uses a slightly different absolute solar $g$-band magnitude of 5.144 \citep[Table 1 of][]{Into2013}.
Like Equation~\ref{eq:gr}, this relation also assumes an exponentially decaying SFH; however, it was determined from theoretical models, rather than semi-empirically from observed photometry. 

The majority of color-$M_\star/L$ relations in the literature, including Equations~\ref{eq:gr} and \ref{eq:gr_ip} above, were calibrated using samples of relatively massive ($M_\star>10^{8-9}~\rm{M_{\odot}}$) galaxies. 
We therefore also consider empirical relations that are specifically based on low-mass galaxies.

\citet{Herrmann2016} measured color-$M_\star/L$ relations for 34 nearby dIrr galaxies from Local Irregulars That Trace Luminosity Extremes, The H I Nearby Galaxy Survey \citep[LITTLE THINGS;][]{Hunter2012}. 
We consider their $g-r$ relation:
\begin{equation}
    \log(M_{\star}/\rm{M_{\odot}}) = 1.731 + 0.894(g-r)_0 - 0.4M_{g,0}.
\end{equation}
Stellar mass density profiles for these galaxies were measured by modeling galaxy SEDs with a multi-component stellar population model.

\citet{Klein2024} proposed a color-$M_\star/L$ relation based on low-mass galaxies in the FIRE-2 simulations (see Section~\ref{sec:sims}):
\begin{equation}
\label{eq:klein24}
    \log(M_{\star}/\rm{M_{\odot}}) = 1.607 + 1.570(g-r)_0 - 0.4M_{g,0}.
\end{equation}
This relation is based on 20 simulated galaxies with stellar masses between $10^{5-10}~\rm{M_{\odot}}$.
\citet{Klein2024} used the \texttt{FIRE\_studio} code \citep{Gurvich2022,Hopkins2005} to create post-processed images via ray projection from the star particles, then measured synthetic photometry from these mock images.

Finally, \citet{Du2020} did not develop a new photometric calibration, but instead found corrections for literature color-$M_\star/L$ relations in order to make them internally self-consistent when using different photometric bands \citep{McGaugh2014}.
They used a sample of low surface brightness galaxies to empirically correct several literature relations, including both \citet{Bell2003} and \citet{Into2013}, listed respectively:
\begin{align}
    \log(M_{\star}/\rm{M_{\odot}}) & = 1.461 + 1.097(g-r)_0 - 0.4M_{r,0} \\
    \log(M_{\star}/\rm{M_{\odot}}) & = 1.254 + 1.530(g-r)_0 - 0.4M_{r,0}
\end{align}

\subsubsection{Infrared (WISE) $M_{\star}/L$ calibrations}
\cite{Jarrett2023} computed updated $M_{\star}/L$ calibrations based on near-infrared photometry from the Wide-field Infrared Survey Explorer (WISE).
They report two stellar mass calibrations, the first based on W1 (3.4$\mu$m) flux and the second based on W1$-$W2 (3.4$\mu$m-4.6$\mu$m) color:
\begin{align}
    \log(M_{\star}/\rm{M_{\odot}}) & = -12.62 + 5.00\log(L_{\mathrm{W1},0}) \nonumber \\
    & - 0.44\log(L_{\mathrm{W1},0})^2 + 0.016\log(L_{\mathrm{W1},0})^3 \label{eq:w1}\\
    \log(M_{\star}/\rm{M_{\odot}}) & = \log(L_{\mathrm{W1},0}) - 0.376 - 1.053(\mathrm{W1} - \mathrm{W2})_0 \label{eq:w12}
\end{align}
Here, $L_{\mathrm{W1},0}=10^{-0.4(M_{\mathrm{W1},0}-M_{\mathrm{W1},\odot})}$, where the W1 in-band solar value is $M_{\mathrm{W1},\odot}=3.24$ mag \citep{Jarrett2013}.
In Equations~\ref{eq:w1} and \ref{eq:w12}, all WISE magnitudes and colors are in Vega magnitudes \citep[to convert from AB to Vega, we follow the prescriptions in Table 1 of][]{Jarrett2011} and are $k$-corrected following Equations A1 and A2 in \citet{Jarrett2023}.
These calibrations were based on galaxies in the Galaxy and Mass Assembly (GAMA) survey with WISE fluxes.
This sample contained galaxies with stellar masses down to $\sim10^{6.5}~\rm{M_{\odot}}$, but is likely incomplete below $\sim10^{8}~\rm{M_{\odot}}$.
The GAMA DR4 stellar masses \citep{Driver2022} were computed by fitting a grid of \cite{BruzualCharlot2003} SPS models \citep[for more details, see][]{Taylor2011} and assumed exponential SFHs, uniform metallicities, and \cite{Calzetti2000} dust attenuation.

\subsection{Prospector}
\label{sec:prospector}

The empirical relations considered in Section~\ref{sec:fluxcalibs} use only a few photometric bands to estimate stellar masses.
Fitting the multi-wavelength SED of a galaxy takes advantage of photometric information across a much broader wavelength range and is frequently considered the ``gold standard'' for obtaining galaxy stellar masses from integrated light.
Prospector \citep{Leja2017,Leja2019,Johnson2021} is a Bayesian inference code that estimates galaxy properties by forward modeling galaxy SEDs with the Flexible Stellar Population Synthesis package \citep[FSPS;][]{Conroy2009,Conroy2010}.

As we describe in Section~\ref{sec:models}, we also use FSPS to produce mock observations of simulated galaxies.
Using Prospector to fit these mock observations is therefore an ``apples-to-apples'' comparison that directly probes the capabilities of SED fitting rather than the effect of other systematic uncertainties (e.g., stellar model libraries, SED fitting algorithm).
We defer tests of these other systematic effects, caused by variations among different SED fitting codes, to future work.

\section{Simulated galaxies} 
\label{sec:models}
We test stellar mass calibrations on mock observations of simulated galaxies, for which the ``true'' stellar masses are known.
Here we describe the models used and their major assumptions, as well as the method used to produce mock observations from the simulated galaxies.

\subsection{Galaxy models}
\label{sec:sims}

\subsubsection{GRUMPY}
The simplest model we consider is GRUMPY \citep[Galaxy formation with
RegUlator Model in PYthon;][]{Kravtsov2022}, a semi-analytic model of the ``regulator'' type. In this model, a low-mass galaxy is simulated using a system of differential equations that track the mass conservation of different baryonic components.
The key input to the model is an underlying dark matter halo accretion history, which is taken from the halo mass accretion histories from the ELVIS high-resolution simulation suite \citep{Garrison-Kimmel2014}. 

The gas inflow rate is assumed to be proportional to the halo accretion rate with additional factors describing suppression due to UV heating (i.e., from reionization) and the formation of a hot gaseous halo. The gas outflow rate is assumed to be proportional to the star formation rate (SFR) with the constant of proportionality (the mass-loading factor) parametrized as a function of stellar mass.    
The SFH is set by the star formation rate, which assumes a constant molecular hydrogen gas depletion time and instantaneous recycling. Only in-situ formed stars are considered in this model; stellar contribution from galaxy mergers is not accounted for in the model. 
The production and removal of heavy element abundances in the galaxy's ISM are also parameterized, providing a chemical evolution history for the galaxy.
Finally, further stochasticity is added to the SFH by introducing a correlated random perturbation in SFR relative to the mean $M_\star - \mathrm{SFR}$ relation at each timestep \citep{Pan2023}.

In total, we obtain 69 galaxy SFHs and chemical enrichment histories from GRUMPY (with boosted stochasticity).
This sample covers a stellar mass range of $10^{4-11}~\rm{M_{\odot}}$.

\subsubsection{UniverseMachine-SAGA}
The next model we consider is the UniverseMachine \citep{Behroozi2019}, an empirical galaxy-halo connection model.
This model assumes that galaxy SFRs depend on host halo properties (halo mass, assembly history, and redshift).
It then fits this relationship by predicting galaxy observables (e.g., stellar mass functions, quenched fractions) from a dark-matter-only simulation, and iteratively comparing the predictions with observations over a wide range of galaxy masses and redshifts simultaneously.
\cite{Wang2021} extended the DR1 of the UniverseMachine model\footnote{\url{https://bitbucket.org/pbehroozi/universemachine/src/main/}} to lower-mass galaxies by applying it to a joint set of 45 dark-matter-only zoom-in simulations of isolated Milky Way–mass halos from the Symphony compilation~\citep{Nadler2023}.

In this work, we use SFHs predicted by the updated version of UniverseMachine, UM-SAGA~\citep{Wang2024}, which is constrained by SAGA satellites~\citep{Geha2017,Mao2021,Mao2024,Geha2024} and SDSS isolated galaxies~\citep{Geha2012} down to $M_{\star}\gtrsim 10^{7}\rm{M_{\odot}}$. UM-SAGA provides a better match to observed low-mass galaxies properties than \textsc{UniverseMachine} DR1, and is particularly better at predicting the quenched fraction of low-mass galaxies.  We apply UM-SAGA to one of the 45 Symphony Milky Way-mass halos and obtain 43 satellite galaxy SFHs.
Here, we define a ``satellite'' galaxy in UM-SAGA as any subhalo within 5 virial radii of the host, with a virial mass of at least $1.2\times10^8~\rm{M_\odot}$ (corresponding to 300 particle masses).

\begin{figure}[t!]
    \centering
    \epsscale{1.15}
    \plotone{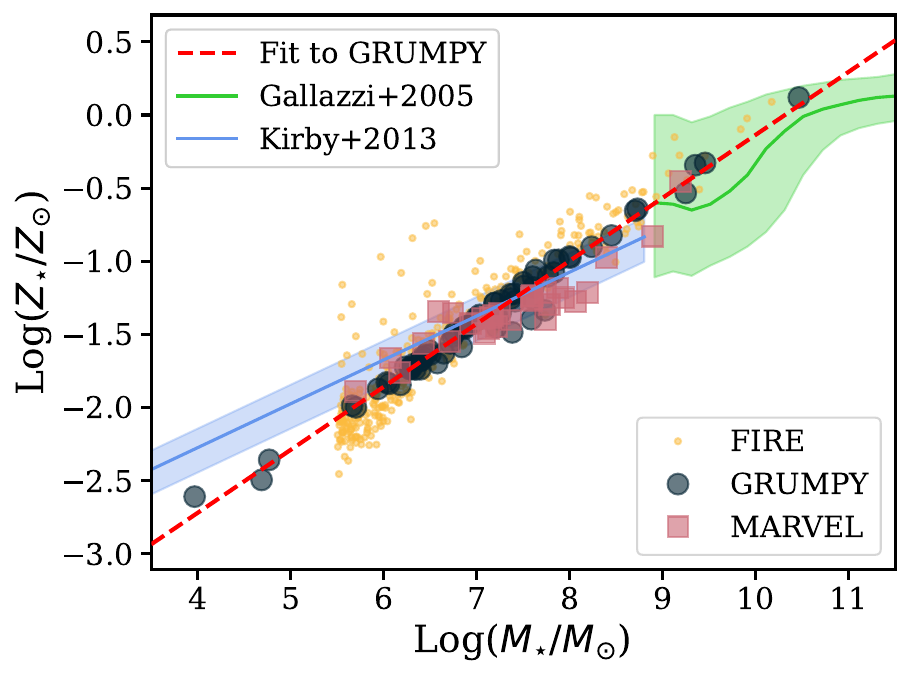}
    \caption{Relationships between stellar mass and stellar metallicity for the simulations that track chemical enrichment: GRUMPY (dark purple points), MARVEL (pink points), and FIRE-2 (yellow points). These models have roughly consistent mass--metallicity relationships, suggesting that any differences between their derived stellar masses are not due to metallicity.
    The red dashed line indicates the best linear fit to the GRUMPY data; as described in Section~\ref{sec:sims}, this fit is used to compute chemical enrichment histories for the UniverseMachine simulated galaxies.
    The observed stellar mass--metallicity relations from \citet{Kirby2013} (blue solid line indicates the best fit to Local Group dwarf galaxies; shaded regions indicate the rms about the best-fit line) and \citet{Gallazzi2005} (green solid line indicates the running median; shaded regions indicate the 16th and 84th percentiles) are also shown for comparison.
    }
    \label{fig:mz_sim}
\end{figure}

Of these 43 simulated galaxies, 35 have stellar masses $<10^{7}~\rm{M_{\odot}}$ where the quenched fraction is not well constrained by observations and may lead to additional uncertainties in the SFHs; we include these galaxies in all plots for illustrative purposes but do not include them in our analysis.\footnote{For completeness, we test the effect of including and excluding these poorly constrained UM-SAGA galaxies in our analysis, and we find that excluding them does not change any of our conclusions.}
This model does not track chemical enrichment, so we produce chemical enrichment histories by applying a mass--metallicity relation to the SFHs. 
For consistency, we obtain this mass--metallicity relation by fitting a simple linear relation\footnote{This and all other fits in this work are, unless otherwise specified, computed with least-squares minimization using a Levenberg-Marquardt algorithm. We bootstrap all fits by performing each fit $N=1000$ times, sampling with replacement on each iteration; the reported best-fit parameters are the median estimates.} to the GRUMPY models (Figure~\ref{fig:mz_sim}):
\begin{equation}
    \label{eq:mz}
    \log\left(\frac{Z_{\star}}{\rm{Z_{\odot}}}\right) = 0.43\log\left(\frac{M_{\star}}{\rm{M_{\odot}}}\right) - 6.24
\end{equation}

\subsubsection{MARVELous Dwarfs}
Finally, we use simulated galaxies from two different cosmological baryonic zoom-in simulations.
The first is the MARVELous Dwarfs \citep{Bellovary2021, Munshi2021, Christensen2024, azartash2024}, a suite of high-resolution zoom-in simulations that use the N-body code ChaNGA \citep{Menon2015}. 
The MARVELous Dwarfs (hereafter abbreviated as MARVEL) include four simulated volumes of low-mass galaxies in isolated environments using a WMAP3 cosmology \citep{Spergel07}. 
These simulations implement gas, initial star, and dark matter particle masses of 1410 $\rm{M_{\odot}}$, 420 $\rm{M_{\odot}}$, and 6650 $\rm{M_{\odot}}$, respectively, and have a force resolution of 60 pc. The high spatial resolution and small particle mass allow MARVEL galaxies to be resolved down to $M_*\approx3\times10^3\rm{M_{\odot}}$. 
MARVEL simulates a number of baryonic processes, including: star formation and gas cooling \citep{Christensen12}, metal line cooling and metal diffusion \citep{Shen10}, non-equilibrium formation and destruction of molecular hydrogen, photoionization and heating from a cosmological UV field background \citep{Haardt12}, and a ``blastwave'' model of supernova feedback \citep{Stinson06} that temporarily shuts off gas cooling during the momentum-conserving ``snowplow phase'' \citep{McKee77}.
The combined processes emulate the energy deposited within the interstellar medium by all processes related to young stars, including UV radiation \citep[see][]{Wise12,Agertz13}. 
The Amiga Halo Finder \citep[AHF;][]{AHF} is applied to MARVEL to identify dark matter halos, subhalos, and the baryonic content within. 
We obtain 30 SFHs and chemical enrichment histories of satellite and field galaxies from MARVEL, with a stellar mass range of $10^{5.7}-10^{9.2}~\rm{M_{\odot}}$.

\subsubsection{FIRE-2}
We also consider the Latte \citep[introduced in][]{Wetzel2016} and ELVIS on FIRE \citep[introduced in][]{Garrison-Kimmel2019a} suites from the FIRE-2 cosmological baryonic zoom-in simulations, part of the Feedback In Realistic Environments project \citep[public data release introduced in][]{Wetzel2023}.
The FIRE-2 simulations are run using Gizmo, a Lagrangian Mesh-less Finite Mass (MFM) hydrodynamics code \citep{Hopkins2015}, and the FIRE-2 physics model \citep{Hopkins2018}.
FIRE-2 models the dense, multiphase interstellar medium
(ISM) in galaxies and incorporates physically motivated, metallicity-dependent radiative heating and cooling processes for gas. These include free-free, photoionization and recombination, Compton, photo-electric and dust collisional, cosmic ray, molecular, metal-line, and fine structure processes. The model tracks $11$ element species (H, He, C, N, O, Ne, Mg, Si, S, Ca, Fe) across a temperature range of $10$ -- $10^{10}$K. FIRE-2 also includes the following time-resolved stellar feedback processes: core-collapse and white dwarf (Type Ia) supernovae, continuous mass loss, radiation pressure, photoionization, and photo-electric heating. 

In this analysis, we use a sample of $362$ low-mass galaxies around MW/M31-mass galaxies from the $z=0$ snapshots from the Latte and ELVIS on FIRE suites, and obtain their SFHs and chemical enrichment histories. The simulations have a dark matter mass resolution of $2 - 3.5 \times 10^4 \,\rm{M_{\odot}}$, and gas and star mass resolution of $3500 - 7100 \, \rm{M_{\odot}}$, which allows for well-resolved galaxies down to $M_* \approx 10^5 \, \rm{M_{\odot}}$. Identifying dark matter halos, and subhalos is done via the ROCKSTAR halo finder \citep{Behroozi2013}.
Importantly, this sample of FIRE-2 galaxies has been benchmarked against the stellar mass
functions, radial distance distributions, and star-formation histories
of low-mass galaxies in the LG \citep{Wetzel2016, Garrison-Kimmel2019a, Garrison-Kimmel2019b, Samuel2020, Samuel2021}.

\vspace{1em}
In total, our sample consists of 469 simulated galaxies\footnote{This number excludes the $<10^7~\rm{M_{\odot}}$ UM-SAGA galaxies with poorly constrained SFHs.} with a mass range of $M_{\star}=10^{4.0-10.8}~\rm{M_{\odot}}$.
The majority of our sample (403 galaxies) are low-mass ``dwarf'' galaxies, with $M_{\star}<10^8~\rm{M_{\odot}}$.
The remaining 66 galaxies constitute a relatively small sample that is likely incomplete at the highest masses ($\gtrsim10^{9.5}~\rm{M_{\odot}}$); however, this high-mass sample still provides a useful comparison when testing different mass measurement techniques.

While it is not possible to determine whether our simulated galaxies are truly representative of real low-mass galaxies, we can at least check whether the models produce scaling relations that agree with each other and with observations.
As shown in Figure~\ref{fig:mz_sim}, both the FIRE-2 and MARVEL galaxies follow roughly the same mass--metallicity relation as the GRUMPY galaxies (and by extension the UM-SAGA galaxies, which by definition follow the red dotted line in Figure~\ref{fig:mz_sim}, given by Equation~\ref{eq:mz}).
Figure~\ref{fig:mz_sim} also shows the observed stellar mass--metallicity relations measured by \citet{Kirby2013} and \citet{Gallazzi2005}. 
The mass--metallicity relation of our sample is somewhat steeper than that measured for Local Group dwarf galaxies by \citet{Kirby2013}, who measured a slope of 0.30 compared to our slope of 0.43, but the two relations agree (within the rms about the \citealt{Kirby2013} best-fit line) in the mass range $10^{6-8}~\rm{M_{\odot}}$, which spans most of our low-mass galaxy sample.
Where our sample overlaps with the higher mass range of the \citet{Gallazzi2005} mass--metallicity relation, our simulated galaxies and simple linear fit are consistent with the \citet{Gallazzi2005} relation within 1$\sigma$.

\subsection{Mock observations}
\label{sec:mockobs}

To produce mock observations of these simulated galaxies, we use the FSPS package \citep{Conroy2009,Conroy2010}\footnote{In particular, we use Python-FSPS, the Python binding of FSPS developed by \citet{pyfsps}, which can be found at \url{https://dfm.io/python-fsps/current/}.}.
FSPS takes galaxy SFHs and chemical enrichment histories as input and generates spectra and photometry.
For all mock galaxies, we use MIST isochrones \citep{Choi2016} and the MILES spectral library \citep{FalconBarroso2011,Vazdekis2010,Vazdekis2015} to model stellar populations and spectra.
We assume a \citet{Calzetti2000} attenuation curve, normalized by setting the optical depth at 5500\AA\ to $\tau_{V}=0.2$.
We include dust emission using the default FSPS parameters.
To further mimic realistic conditions, we place all simulated galaxies at a redshift of $z=0.01$, consistent with the distances probed by ongoing and upcoming surveys of the nearby universe \citep[e.g., the DESI LOW-$Z$ survey aims to target low-mass galaxies out to $z<0.03$;][]{Darragh-Ford2023}.\footnote{We note that this is not strictly necessary, since many of the tests described in this work assume that the redshift is exactly known. However, we plan to test the effect of unknown redshift on stellar mass measurements in future works.}
We note that when SFHs are given as tabular input, the current version of FSPS cannot include nebular continuum or line emission in the output photometry; we defer testing the effect of nebular emission to a later work.

\begin{deluxetable}{llc}
\tablecolumns{3} 
\tablecaption{Photometric filters used to produce mock galaxy SEDs. \label{tab:bands}} 
\tablehead{ 
\colhead{Instrument} & \colhead{Filter} & \colhead{$\lambda_{\mathrm{eff}}$ (\AA)}
}
\startdata
GALEX & Far-UV & 1549 \\
 & Near-UV & 2304 \\
SDSS & $g$ & 4640 \\
 & $r$ & 6122 \\
 & $z$ & 8897 \\
WISE & W1 & 33680 \\
 & W2 & 46180 \\
\enddata
\end{deluxetable}

\begin{figure}[t!]
    \centering
    \epsscale{1.17}
    \plotone{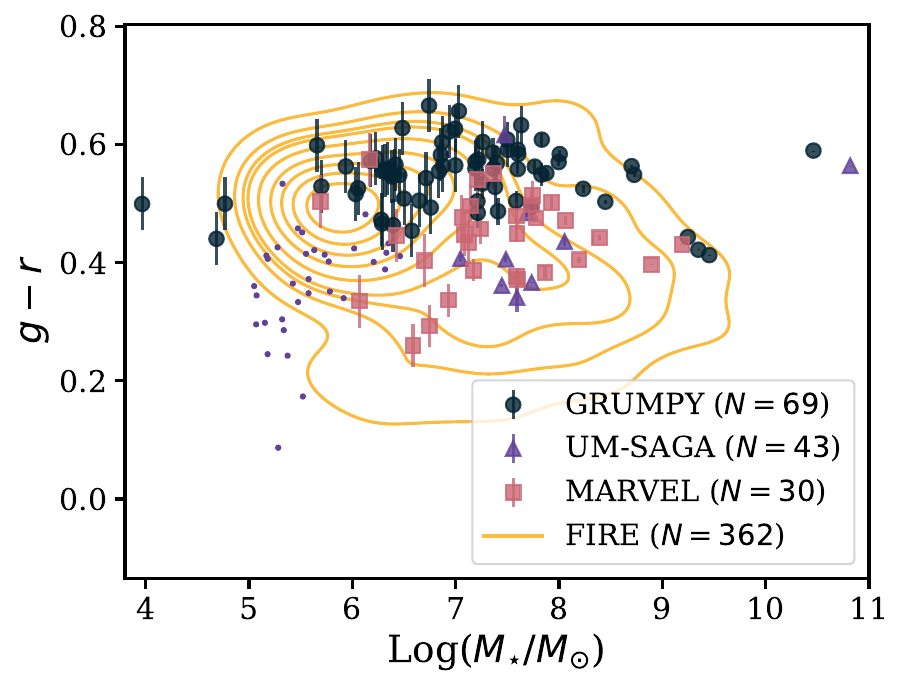}
    \caption{The $g-r$ color as a function of stellar mass for the GRUMPY (dark purple circles), UM-SAGA (purple triangles), MARVEL (pink squares), and FIRE-2 (yellow contours) models. UM-SAGA SFHs are not well constrained below $\log(M_\star/\rm{M_{\odot}})\sim7$, so these galaxies are marked with small points; for clarity, FIRE-2 galaxies are represented as density contours rather than points.}
    \label{fig:gr_sim}
\end{figure}

Using the input SFH and chemical evolution history of a galaxy, FSPS calculates the surviving stellar mass of the galaxy.
We use this as ``true'' stellar mass of the galaxy, rather than the raw output mass measured from GRUMPY, UM-SAGA, MARVEL, or FIRE.
This is because the four models assume a variety of different stellar physics models, including a range of mass loss rates, and these differences can lead to systematic discrepancies in galaxy $M_\star$ measurements.
We therefore use FSPS to self-consistently convert between the \emph{formed} SFH and \emph{surviving} stellar mass, using the same conversion for all four models.
With the MIST isochrones \citep{Choi2016}, this corresponds to a typical difference of $\sim40-50\%$ between the formed and surviving $M_\star$.

FSPS generates noiseless photometry in bands ranging from the UV through IR, listed in Table~\ref{tab:bands}.
Our choice of bands is motivated by the existence of wide-area surveys---GALEX in the UV \citep{Martin2005,Bianchi2017}, SDSS in the optical \citep{Blanton2017,Doi2010}, and WISE in the near-IR \citep{Wright2010}---which, combined, provide multi-wavelength photometry for $>700,000$~galaxies in the low-redshift universe \citep{Salim2016}. 
We then add realistic noise to the baseline ``noiseless'' photometry in order to compare directly with observations.
As a reference, we use the third data release of the Satellites Around Galactic Analogs survey \citep[SAGA DR3;][]{Mao2024}, which identified 378 satellite galaxies around 101 Milky Way-mass analogs between 25--40.75 Mpc away.
SAGA used imaging data from several public datasets---in particular, the optical $grz$ photometry came from the DESI Imaging Legacy Surveys \citep{Dey2019}.
To estimate observational noise for each galaxy in our sample, we first use a k-d tree to identify the SAGA satellite with the nearest $r$-band magnitude, $g-r$ color, and $r-z$ color.
For each photometric band $X$, we then assign the photometric uncertainty $\sigma_{X}$ from the nearest-neighbor SAGA satellite to the simulated galaxy, and we perturb the simulated galaxy's noiseless photometry by adding a noise factor $\delta_{X}$ randomly drawn from the Gaussian distribution $\delta_{X}\sim\mathcal{N}(0,\sigma_{X})$. 

\begin{figure}[t!]
    \centering
    \epsscale{1.17}
    \plotone{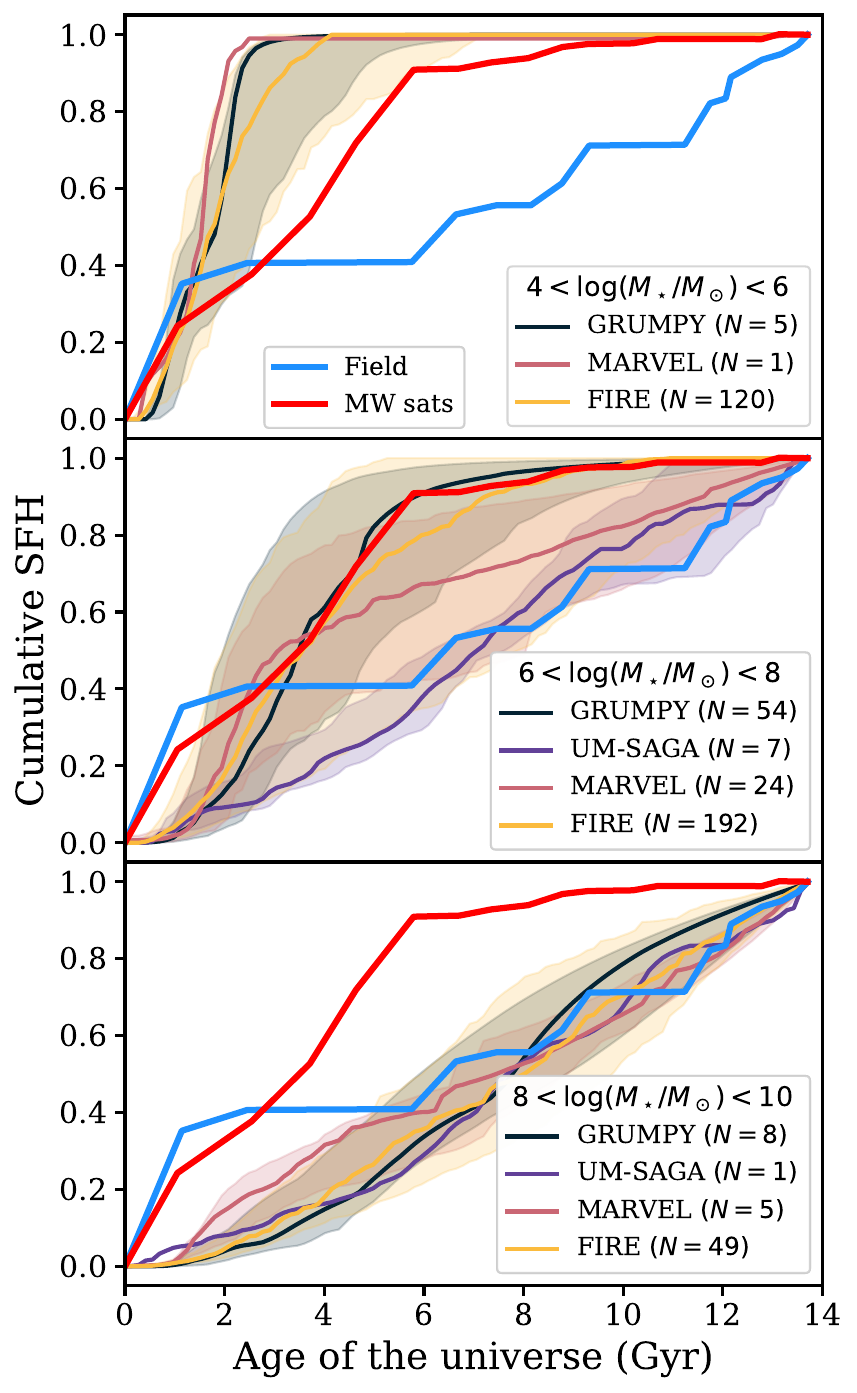}
    \caption{Cumulative SFHs of the simulated galaxies considered in this work. The panels illustrate galaxies of different stellar mass ranges: $10^{4-6}~\rm{M_\odot}$ (top), $10^{6-8}~\rm{M_\odot}$ (middle), and $10^{8-10}~\rm{M_\odot}$ (bottom). UM-SAGA galaxies with $M_\star<10^7~\rm{M_{\odot}}$ are not included in this plot. Solid lines mark the median SFHs of the galaxies from GRUMPY (dark purple), UM-SAGA (purple), MARVEL (pink), and FIRE-2 (yellow). The corresponding filled regions denote the 16th-to-84th percentile spread of the SFHs. For comparison, the median SFHs for field dwarf galaxies (blue) and Milky Way satellites (red) measured from resolved color-magnitude diagrams by \citet{Weisz2014} are shown.}
    \label{fig:sfh_sim}
\end{figure}

\begin{figure}[t!]
    \centering
    \epsscale{1.17}
    \plotone{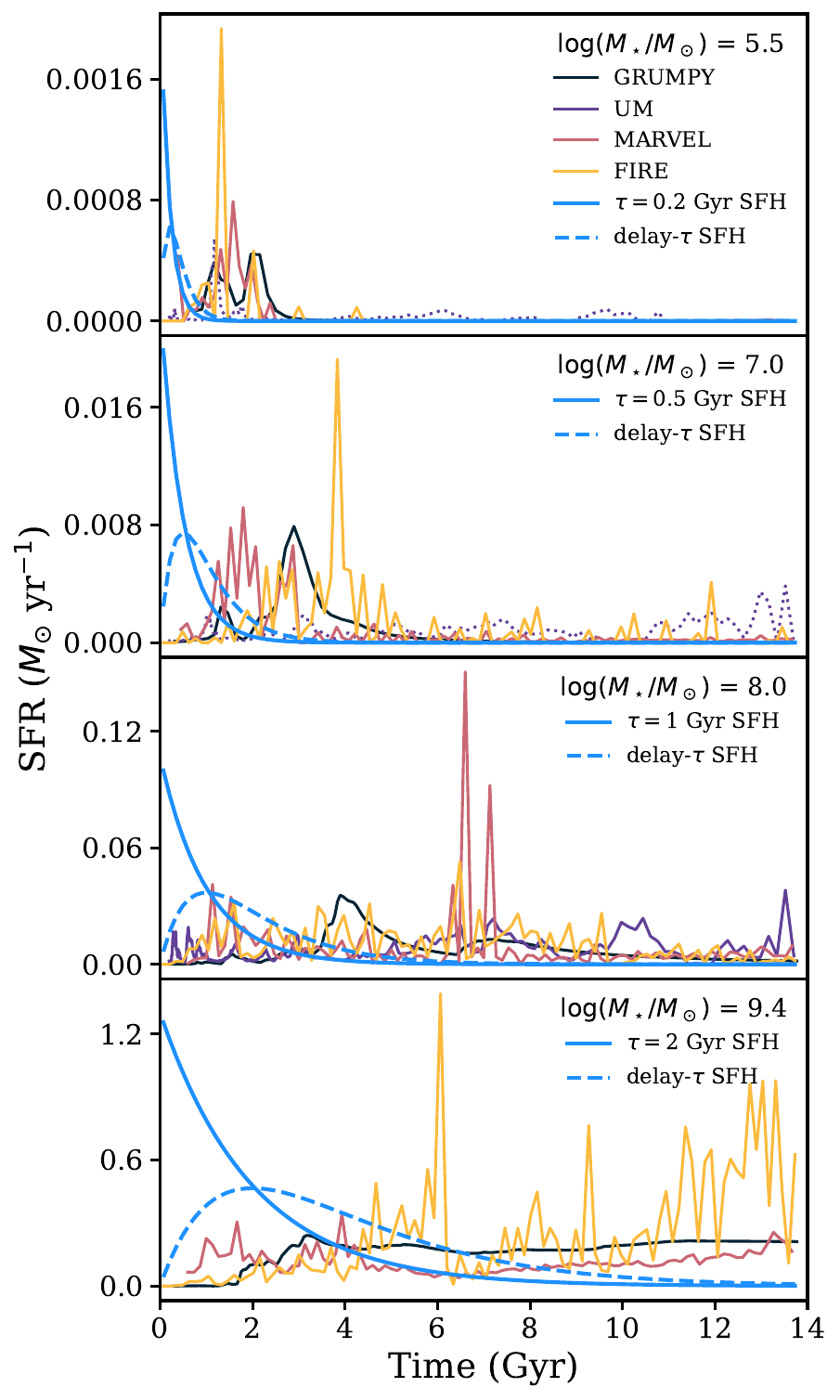}
    \caption{Example SFHs from each of the simulations. The panels illustrate galaxies of different stellar masses: (from top to bottom) $10^{5.5}$, $10^{7.0}$, $10^{8.0}$, and $10^{9.4}~\rm{M_\odot}$. Solid lines mark the median SFHs of the galaxies from GRUMPY (dark purple), UM-SAGA (purple), MARVEL (pink), and FIRE-2 (yellow); UM-SAGA galaxies with $M_\star<10^7~\rm{M_{\odot}}$ are shown as dotted lines. For comparison, exponentially decaying $\tau$ ($\propto e^{-t/\tau}$) and delay-$\tau$ SFHs (Equation~\ref{eq:delaytau}) are shown as blue lines (solid and dashed, respectively).}
    \label{fig:samplesfh}
\end{figure}

Figure~\ref{fig:gr_sim} illustrates the results of this FSPS modeling, showing the distribution of the simulated galaxies in $(g-r)$--mass space.
In general, the galaxies from all four models (particularly when the poorly-constrained $<10^{7}~\rm{M_\odot}$ UM-SAGA galaxies are removed) occupy a similar region of optical color space.
However, each of the models described in Section~\ref{sec:sims} parameterize galaxy evolution differently, which leads to some minor variations among the mock observations.
For example, the GRUMPY galaxies are systematically redder than the UM-SAGA galaxies, particularly in the mass range $10^{6-8}~\rm{M_\odot}$.
This is likely related to differences in the model SFHs, which are plotted in Figures~\ref{fig:sfh_sim} and \ref{fig:samplesfh}. 
As shown in both the middle panel of Figure~\ref{fig:sfh_sim} and the top three panels of Figure~\ref{fig:samplesfh}, the GRUMPY galaxies (dark purple) typically form more stars at earlier times, whereas the UM-SAGA galaxies (purple) have more extended SFHs and higher recent star formation rates, leading to bluer colors.
We revisit potential implications of these differences on stellar mass calibrations in Section~\ref{sec:sim_differences}.

Notably, Figure~\ref{fig:samplesfh} also illustrates stochasticity in the model SFHs, particularly at lower masses.
This is in sharp contrast with exponentially decaying SFH models (blue lines), which are used both to calibrate many of the color-$M_\star/L$ relations described in Section~\ref{sec:calibrations}, and to parametrize SFHs in SED fitting codes.
In the following sections, we will see how this burstiness affects stellar mass estimates of low-mass galaxies.

\begin{figure*}[t!]
    \centering
    \epsscale{1.17}
    \plotone{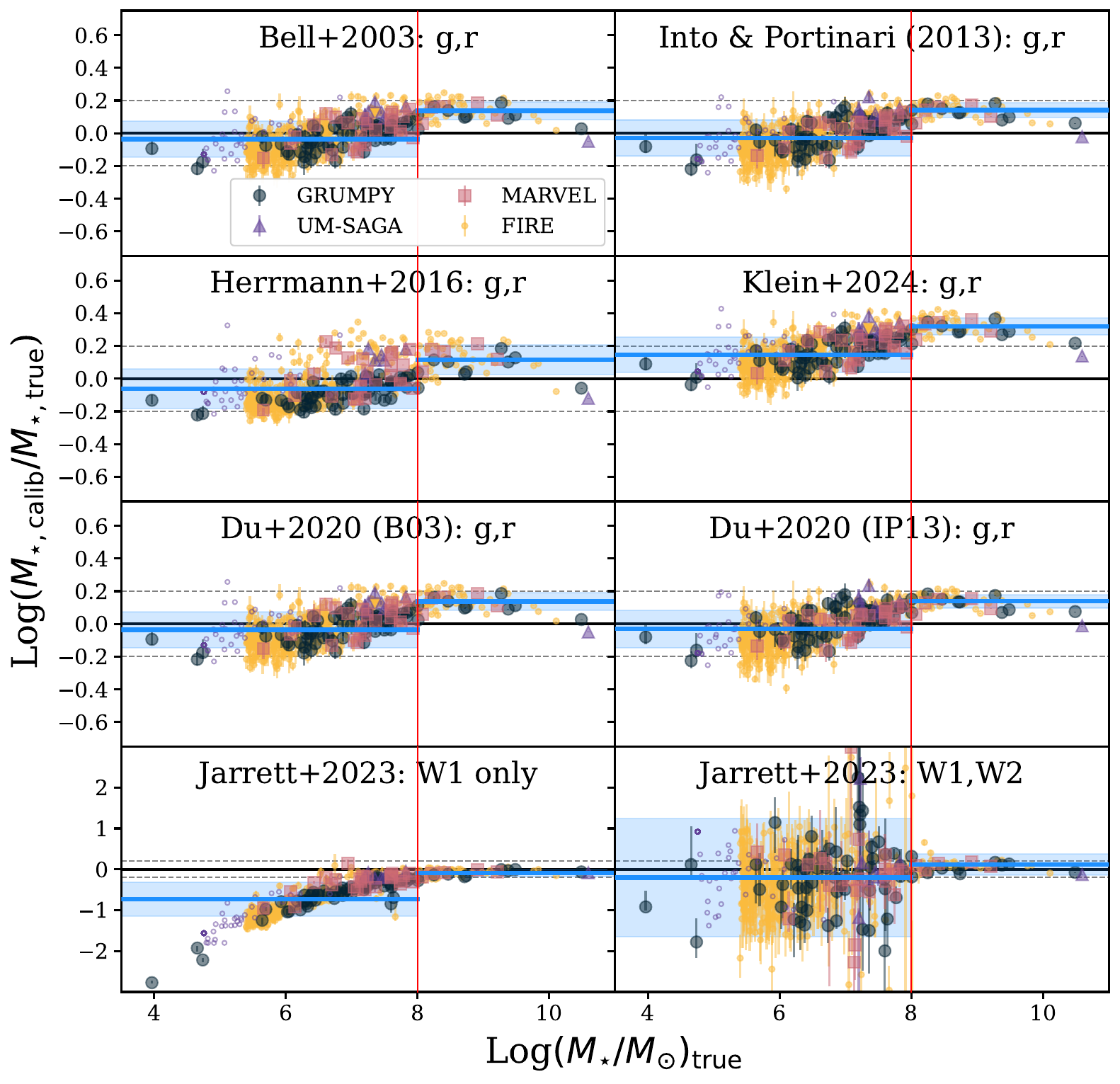}
    \caption{Comparisons between ``true'' stellar masses of model galaxies ($M_{\star,\mathrm{true}}$) and stellar masses measured from mock observations using different empirical photometric calibrations ($M_{\star,\mathrm{calib}}$). 
    Different colors and shapes represent mock observations obtained from different galaxy models. UM-SAGA (purple) galaxies with $\log(M_{\star}/\rm{M_{\odot}})<7$ are marked with small unfilled points to denote that their SFHs are not well constrained. The dashed horizontal gray lines mark $\pm0.2$~dex errors, which represent typical systematic uncertainties that are assumed for stellar mass measurements.
    The horizontal blue lines and shaded bands denote the means and standard deviations of the mass residuals for the low-mass ($M_{\star}<10^{8}~\rm{M_{\odot}}$) and high-mass ($M_{\star}>10^{8}~\rm{M_{\odot}}$ samples, separated by the vertical red line.)
    Note the expanded $y$-axis limits on the bottom two plots.}
    \label{fig:sim_empirical}
\end{figure*}

\begin{deluxetable*}{lllllll}
\tablecolumns{7} 
\tablecaption{Residuals between ``true'' and measured masses, $\mathrm{log}\left(M_{\star,\mathrm{calib}}/M_{\star,\mathrm{true}}\right)$, for different photometric calibrations. The ``Bands'' column lists the various photometric bands used in each calibration, and the bolded band is the reference band (e.g., when calculating $M_\star/L_{g}$, $g$ is the reference band). The values reported are the mean and standard deviation of the residuals for low-mass ($M_{\star,\mathrm{true}} < 10^{8}~\rm{M_{\odot}}$) galaxies. \label{tab:photresiduals}} 
\tablehead{ 
\multicolumn{7}{c}{Photometric calibrations} \\
\colhead{Calibration} & 
\colhead{Bands} &
\colhead{GRUMPY} & 
\colhead{UM-SAGA\tablenotemark{a}} & 
\colhead{MARVEL} &
\colhead{FIRE} & 
\colhead{All models}
}
\startdata
\citet{Bell2003} & $g,\mathbf{r}$ & $-0.02\pm0.08$ & $\nm0.13\pm0.04$ & $\nm0.01\pm0.09$ & $-0.05\pm0.11$ & $-0.04\pm0.11$ \\
\citet{Into2013} & $\mathbf{g},r$ & $\nm0.00\pm0.09$ & $\nm0.13\pm0.04$ & $\nm0.01\pm0.08$ & $-0.04\pm0.11$ & $-0.03\pm0.11$ \\
\citet{Herrmann2016} & $\mathbf{g},r$ & $-0.09\pm0.07$ & $\nm0.11\pm0.06$ & $\nm0.01\pm0.11$ & $-0.07\pm0.13$ & $-0.06\pm0.12$ \\
\citet{Klein2024} & $\mathbf{g},r$ & $\nm0.16\pm0.08$ & $\nm0.31\pm0.04$ & $\nm0.19\pm0.08$ & $\nm0.14\pm0.11$ & $\nm0.15\pm0.11$ \\
\citet{Du2020} \citepalias{Bell2003} & $g,\mathbf{r}$ & $-0.02\pm0.08$ & $\nm0.13\pm0.04$ & $\nm0.01\pm0.09$ & $-0.05\pm0.11$ & $-0.04\pm0.11$ \\
\citet{Du2020} \citepalias{Into2013} & $g,\mathbf{r}$ & $\nm0.01\pm0.10$ & $\nm0.13\pm0.05$ & $\nm0.00\pm0.08$ & $-0.04\pm0.12$ & $-0.03\pm0.11$ \\
\citet{Jarrett2023} & \textbf{W1} & $-0.65\pm0.46$ & $-0.15\pm0.06$ & $-0.32\pm0.23$ & $-0.79\pm0.39$ & $-0.73\pm0.41$ \\
\citet{Jarrett2023} & \textbf{W1}, W2 & $\nm0.08\pm2.11$ & $\nm0.12\pm0.96$ & $\nm0.63\pm3.91$ & $-0.33\pm0.73$ & $-0.20\pm1.44$ \\
This work & $\mathbf{g},r$ & $\nm0.01\pm0.06$ & $\nm0.10\pm0.03$ & $\nm0.02\pm0.07$ & $\nm0.04\pm0.08$ & $\nm0.03\pm0.07$
\vspace{0.5em}
\enddata
\tablenotetext{a}{UM-SAGA galaxies with $\log(M_\star/\rm{M_\odot})<7$ are not included when calculating these summary statistics due to their poorly constrained SFHs.}
\end{deluxetable*}

\section{Results: Stellar mass recovery using photometric calibrations}
\label{sec:results_photcalib}

We now test how well different stellar mass measurement techniques can recover the ``true'' stellar masses from mock observations, beginning with empirical photometric calibrations based on optical and IR fluxes.

\subsection{Literature calibrations}
\label{sec:litcalibs}

As discussed in Section~\ref{sec:fluxcalibs}, many studies in the literature have developed optical-NIR color-$M_\star/L$ relations to estimate galaxy stellar mass, using either samples of observed galaxies \citep[e.g.,][]{Bell2003,Portinari2004,Schombert2006,Taylor2011,Herrmann2016,Du2020} or theoretical SEDs \citep[e.g.,][]{Zibetti2009,Gallazzi2009,Into2013,Roediger2015,Kouroumpatzakis2023,Klein2024}.
Figure~\ref{fig:sim_empirical} illustrates how well several of these calibrations work on simulated low-mass ($M_{\star,\mathrm{true}}<10^8~\rm{M_{\odot}}$) galaxies. 
Table~\ref{tab:photresiduals} summarizes their performance, listing the mean and standard deviation of the residuals $\log(M_{\star,\mathrm{calib}}/M_{\star,\mathrm{true}})$ for low-mass galaxies from each model discussed in Section~\ref{sec:models}, as well as for all models combined.

The optical color-$M_{\star}/L$ relations (top three rows of Figure~\ref{fig:sim_empirical}) show two primary features.
First, most of the optical calibrations are able to recover $M_\star$ of low-mass galaxies ($<10^8~\rm{M_\odot}$, left of the vertical red lines in the figure) within $\sim0.2$~dex of the true values (dashed horizontal gray lines in the figure).
In this low-mass regime, nearly all optical calibrations have average residuals that are consistent with zero within uncertainties, suggesting that they are statistically unbiased.

The main exception is the calibration by \citet{Klein2024} (right column, second from top panel in Figure~\ref{fig:sim_empirical}), which appears to overestimate $M_\star$ for low-mass galaxies by $0.15\pm0.11$~dex on average. 
This is due to different definitions of ``true'' stellar mass: \citet{Klein2024} determined ``true'' stellar masses from counting star particles in simulated low-mass galaxies in the FIRE-2 simulation, while we use the surviving stellar mass calculated by FSPS from the input SFH (Section~\ref{sec:mockobs}).
\citet{Klein2024} have already showed that their particle-counting method produces stellar masses that are on average $\sim0.17$~dex higher than those estimated from mock observations.
This difference is consistent with the discrepancy observed in our results. 
They attribute this to the assumption of a spatially constant $M_\star/L$ ratio for each galaxy, which may underestimate the mass of star particles in galaxy outskirts.
Such spatial effects are outside the scope of this work, which primarily concerns integrated light measurements.

Even after accounting for this discrepancy, we find that \emph{all} optical relations (top three rows of Figure~\ref{fig:sim_empirical}) show clear systematic trends in the residuals $\log(M_{\star,\mathrm{calib}}/M_{\star,\mathrm{true}})$ as a function of stellar mass.
As shown by the discontinuous blue horizontal lines in the figure, the residuals for the low-mass galaxies ($<10^8~\rm{M_\odot}$, left of the red vertical lines) are consistently $\gtrsim0.15$~dex lower than the residuals for high-mass galaxies.
This is likely related to the SFH: nearly all of the studies discussed in Section~\ref{sec:fluxcalibs} \citep[e.g.,][]{Bell2003,Into2013} use exponentially decaying $\propto e^{-t/\tau}$ (sometimes called ``$\tau$ model'') SFHs to derive optical color-$M_\star/L$ relations, which have been shown to underestimate stellar masses \citep{Li2022}. 
As Figure~\ref{fig:samplesfh} shows, these smooth parametric SFHs (blue lines) are poor fits to the bursty SFHs of the simulated galaxies in our sample; this problem is exacerbated at $\sim10^9~\rm{M_\odot}$, since many of the massive galaxies in our sample have recent bursts of star formation (bottom two panels).

The literature calibrations based on near-IR WISE bands perform more poorly than the optical calibrations (bottom two panels of Figure~\ref{fig:sim_empirical}).
The \citet{Jarrett2023} W1 calibration (bottom left panel) has a pronounced downward trend in $\log(M_{\star,\mathrm{calib}}/M_{\star,\mathrm{true}})$ as $\log{M_{\star,\mathrm{true}}}$ decreases below $10^{8}~M_{\odot}$, which causes it to underestimate $M_\star$ of low-mass galaxies by $0.73$~dex on average (blue horizontal line on the left side of the plot).
Meanwhile, the W1$-$W2 calibration (bottom right panel) produces $M_\star$ residuals with a large $1\sigma$ dispersion of $1.46$~dex for low-mass galaxies (blue shaded region on the left side of the plot).

These trends are likely due to a combination of factors.
The \citet{Jarrett2023} relation based on W1 is a quadratic function of $\log(L_{\mathrm{W1}})$, which can produce strong systematic biases when extrapolated.
The \citet{Jarrett2023} W1$-$W2 calibration, on the other hand, shows dramatic scatter in mass residuals that is almost certainly caused by uncertainties in the W2 band.
For the low-mass ($<10^8~M_\odot$) galaxies in our sample, the assigned uncertainty in W2 magnitude (calculated by finding the nearest-neighbor SAGA galaxy in optical colors; see Section~\ref{sec:mockobs}) is on average 6.2 times larger than the uncertainty in W1 magnitude.
For high-mass galaxies, the W2 magnitude is only 3.3 times more uncertain than the W1 band, which explains why the scatter in the \citet{Jarrett2023} W1$-$W2 relation decreases significantly above $10^8~M_\odot$.
Our results demonstrate that the uncertainties in WISE W2 fluxes make W2 unhelpful in constraining stellar masses. 
However, we are not advocating the exclusion of near-IR fluxes from color--$M_\star/L$ relations; a large sample of more precise near-IR measurements from, e.g., JWST could be used to develop a more reliable stellar mass calibration.

We also note that for all galaxies in our sample, the mock photometric data were produced assuming a single dust attenuation model: the \citet{Calzetti2000} attenuation law with $\tau_V=0.2$.
Changing the dust model may impact the stellar masses recovered by empirical calibrations.
We test this by varying the dust attenuation law in two different ways: we either change the \emph{form} of the dust law by using an SMC-like attenuation law from \citet{Gordon2003}, or we change the total \emph{amount} of dust by increasing the normalization of the \citet{Calzetti2000} law from $\tau_V=0.2$ to up to $\tau_V=1.8$.
In both cases, we find that while altering the dust model may affect the average stellar mass residuals $\log(M_{\star,\mathrm{calib}}/M_{\star,\mathrm{true}})$ (i.e., the blue horizontal lines in Figure~\ref{fig:sim_empirical}), these changes are typically smaller than the 1$\sigma$ dispersion in the residuals.
Furthermore, changing the dust law has the same effect on both low- and high-mass galaxies, so the systematic trends in $\log(M_{\star,\mathrm{calib}}/M_{\star,\mathrm{true}})$ as a function of stellar mass remain unchanged.

\begin{figure}
    \centering
    \plotone{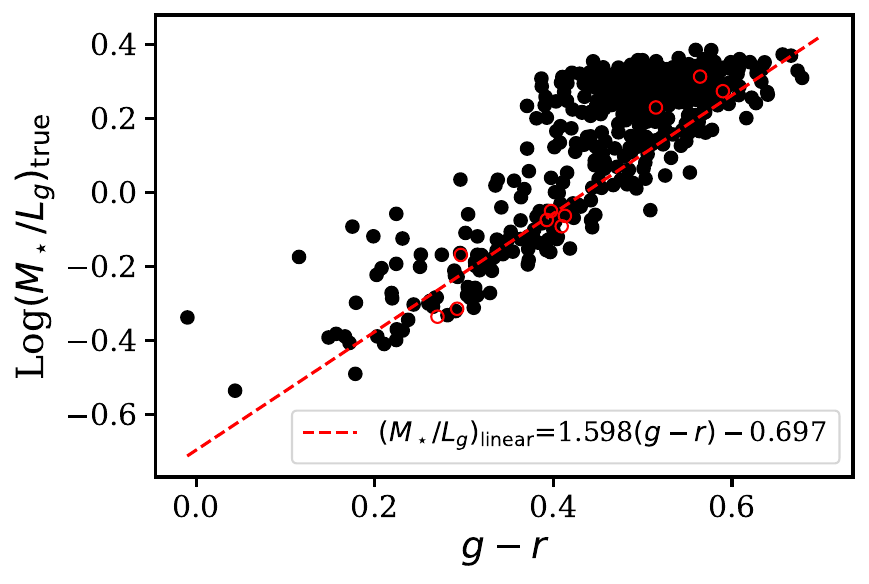}
    \plotone{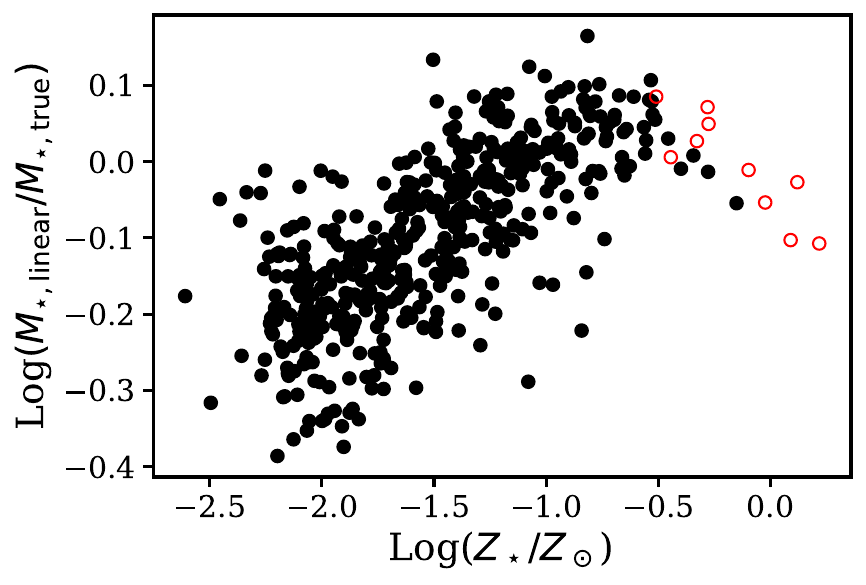}
    \includegraphics[scale=0.45]{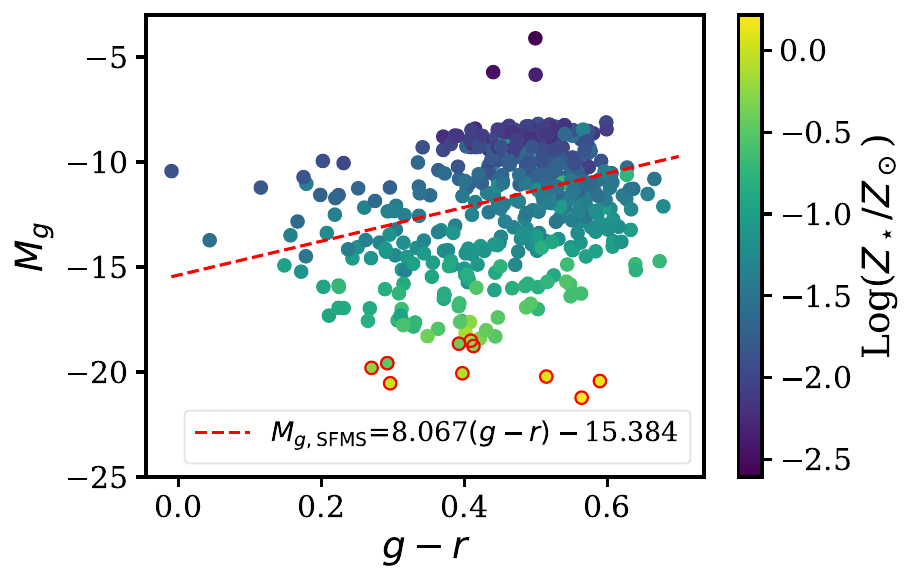}
    \includegraphics[scale=0.45]{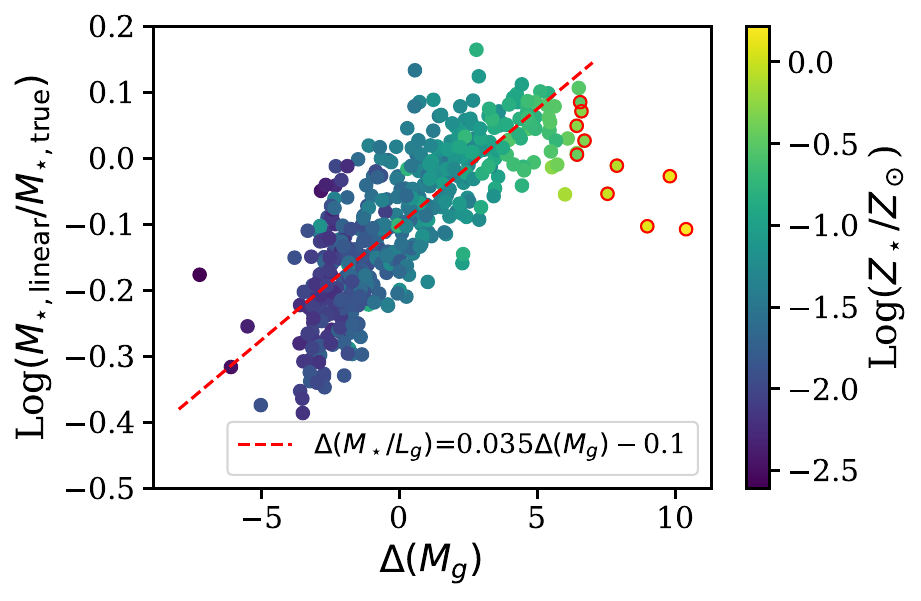}
    \caption{Steps used to calculate the updated mass calibration (Equation~\ref{eq:ourcalib_opt}). Top: initial linear fit (red dashed line) to stellar mass-to-light ratio as a function of optical color. Second from top: stellar mass residuals from initial linear fit as a function of stellar metallicity (known from simulations). Third from top: color-magnitude diagram, showing that distance $\Delta(M_g)$ from the best-fit ``star-forming main sequence'' (red dashed line) can be used as a proxy for metallicity (colorbar). Bottom: stellar mass residuals from initial linear fit as a function of $\Delta(M_g)$. The brightest galaxies ($M_{g,0}<-18.5$; red open circles in all panels) are excluded from the final calibration.}
    \label{fig:newphotcalib}
\end{figure}

\subsection{Updated optical stellar mass calibration}
As discussed in the previous section, many literature optical and NIR calibrations for $M_\star/L$ appear to be systematically biased as a function of stellar mass.
We can use our mock dataset to define a new $M_\star/L$ calibration based on optical colors that minimizes these biases.
As before, in this section all magnitudes have been $k$-corrected using $g-r$ color \citep{Chilingarian2010}, and an absolute solar $g$-band magnitude of 5.11 is assumed \citep{Willmer2018}.

We begin by attempting to fit $M_\star/L_g$ as a linear function of $g-r$ color, as has previously been done in the literature.
From this best fit, we calculate an initial estimate for stellar mass $M_{\star,\mathrm{linear}}$ (top panel of Figure~\ref{fig:newphotcalib}):
\begin{equation}
    \label{eq:originalfit}
    \log(M_{\star,\mathrm{linear}}/\rm{M_\odot}) = 1.598 + 1.347(g-r) - 0.4M_{g,0}
\end{equation}
Our goal is to identify a correction factor that can improve upon this initial estimate.

As shown in the second panel of Figure~\ref{fig:newphotcalib}, the residuals between the true $M_\star/L_g$ and this initial estimate ($\log(M_{\star,\mathrm{linear}}/L_g)-\log(M_{\star,\mathrm{true}}/L_g)=\log(M_{\star,\mathrm{linear}}/M_{\star,\mathrm{true}})$) appear to be strongly correlated with stellar metallicity, which is known precisely from the simulations.
Notably, the more massive galaxies in our sample ($-20.5<M_{g,0}<-18.5$, corresponding to approximate stellar masses of $10^{9.4}\lesssim M_\star/\rm{M_\odot}\lesssim10^{11}$), marked with red open circles in all panels of Figure~\ref{fig:newphotcalib}, do not follow this trend.
Since these bright galaxies are generally well-fit by the initial mass estimate in Equation~\ref{eq:originalfit}, we do not include them in our stellar mass correction (or in any of the following fits in this section).

Of course, it is not possible to use the stellar metallicity to correct a stellar mass estimate if one does not know the true stellar metallicity of a galaxy. 
To make our correction factor useful to an observer who only has access to broadband photometry, we must find a way to estimate (relative) stellar metallicity using $g$ and $r$ photometry.
To do this, we consider the ``fundamental metallicity relation'' (see \citealt{Maiolino2019} for a recent review), which suggests that at a given stellar mass, gas-phase metallicity is anti-correlated with SFR. 
\citet{Geha2024} find that this relation exists for low-mass galaxies, and other studies have identified an analogous relationship for \emph{stellar} metallicity \citep{Looser2004}.
We can represent this relation with photometric proxies: at a given $g$-band magnitude (stellar mass), galaxies with bluer $g-r$ colors (higher star formation rates) should have lower metallicities.
The third panel of Figure~\ref{fig:newphotcalib} shows that such an anti-correlation does exist; metallicity is a strong function of the distance from the mean relationship between the best-fit ($g-r$)-$M_{g,0}$ relation, which is physically analogous to the ``star formation main sequence'' \citep[the relationship between stellar mass and SFR for star-forming galaxies; e.g.,][]{Brinchmann2004}.
This ``star formation main sequence'' can be parameterized with a linear fit (red dashed line in the third panel):
\begin{equation}
    \label{eq:metallicitycalib}
    M_{g,\mathrm{SFMS}} = 8.067(g-r) - 15.384
\end{equation}

The distance between a galaxy's actual magnitude and the ``star formation main sequence,'' $\Delta(M_g) = M_{g,\mathrm{SFMS}} - M_{g,0}$, can therefore be used as a proxy for stellar metallicity.
As shown in the bottom panel of Figure~\ref{fig:newphotcalib}, the residuals $\log(M_{\star,\mathrm{linear}}/M_{\star,\mathrm{true}})$ are strongly correlated with $\Delta(M_g)$.
We again fit this linear relationship (ignoring the bright $M_{g,0}<-17.75$ galaxies, as discussed above):
\begin{equation}
    \label{eq:Mgcalib}
    \log(M_{\star,\mathrm{linear}}/M_{\star,\mathrm{true}}) = 0.035\Delta(M_g) - 0.1
\end{equation}

\begin{figure}[t!]
    \centering
    \epsscale{1.16}
    \plotone{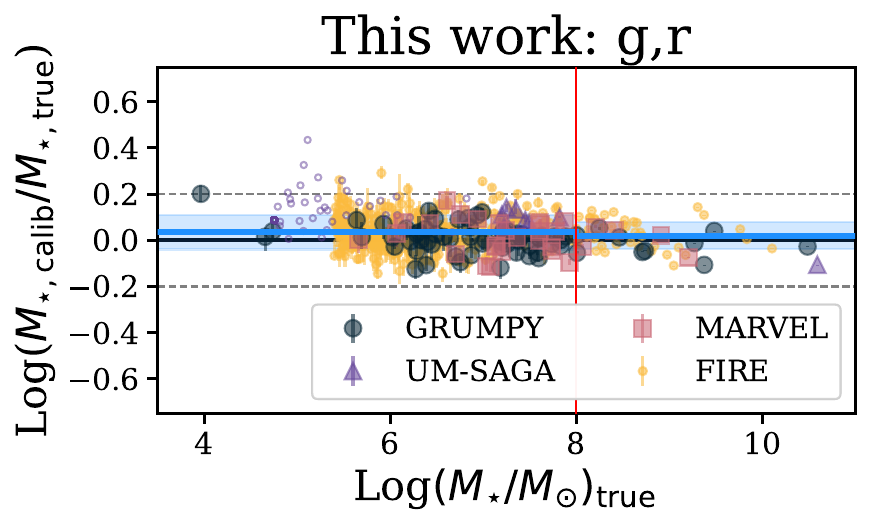}
    \caption{Same as Figure~\ref{fig:sim_empirical}, showing the comparison between the ``true'' stellar masses and stellar masses measured using our updated optical calibration. The dashed horizontal gray lines mark $\pm0.2$~dex errors, the value typically assumed for stellar mass uncertainties. Note the lack of systematic trend in mass residuals compared to Figure~\ref{fig:sim_empirical}; the average residuals of the low-mass and high-mass samples (horizontal blue lines to the left and right of the vertical red dashed line, respectively) are now consistent within standard deviations (horizontal blue shaded regions).}
    \label{fig:ourcalib}
\end{figure}

Combining Equations~\ref{eq:originalfit}-\ref{eq:Mgcalib}, we now have a new calibration for stellar mass based only on $g-r$ color and $g$-band magnitude:
\begin{multline}
    \label{eq:ourcalib_opt}
    \log(M_\star/\rm{M_\odot}) = \\ %1.433(g-r) \\+ 0.00153M_{g,0}^2 - 0.335M_{g,0} + 2.072 %\\ 
    \begin{cases}
    1.986 + 1.315(g-r) - 0.365M_{g,0} & M_{g,0} > -18.5 \\
    1.598 + 1.347(g-r) - 0.4M_{g,0} & M_{g,0} \leq -18.5
    \end{cases}
\end{multline}
The performance of this updated calibration is shown in Figure~\ref{fig:ourcalib}.
This calibration does not appear to have significant systematic residual trends as a function of mass; the galaxies below and above $10^8~\rm{M_\odot}$ have residuals that are consistent within $1\sigma$ dispersion. 

We note some important caveats when using this new calibration.
We caution that given the small size of our simulated sample at low and high masses, this relation is likely to suffer from extrapolation errors outside of the magnitude range $-18.5<M_{g,0}<-5.5$ (corresponding to an approximate mass range of $10^{5.5}\lesssim M_\star/\rm{M_\odot}\lesssim10^{9.4}$).
Furthermore, in our new calibration $M_\star/L$ is no longer a luminosity-independent function.
In principle, a robust $M_\star/L$ calibration for a stellar population should be independent of extensive properties like luminosity or mass.
The implicit luminosity dependence in Equation~\ref{eq:ourcalib_opt} arises from systematic trends in the model SFHs we use: in our sample of model galaxies, low-mass and high-mass have systematically different SFHs (see, e.g., the stochasticity in the low-mass galaxy SFHs shown in Figure~\ref{fig:samplesfh}).

We briefly comment on the impact of the assumed dust attenuation model on our new photometric calibration.
As with the literature calibrations, we find that if we vary the form or normalization of the dust attenuation law in the simulated galaxies, the average stellar mass residuals vary slightly.
However, this effect is again statistically insignificant; regardless of the dust model assumed in the mock photometry, Equation~\ref{eq:ourcalib_opt} produces stellar mass residuals that are approximately constant as a function of stellar mass.

\section{Results: Stellar mass recovery using SED fitting}
\label{sec:results_sed}

We now consider stellar masses estimated from SED fitting.
In a Bayesian framework, the assumptions about prior distributions can have significant effects on the best-fit parameters, and SED fitting is no exception.
In this work, we focus on four primary assumptions that may impact the recovery of $M_{\star}$ for low-mass galaxies: (1) the form of the SFH, (2) the dust model, (3) the IMF, and (4) the choice of photometric bands.
These choices are particularly impactful in the low-mass regime, since low-mass galaxies have stochastic SFHs that may not be described well by classical SFH models and may lead to stochastic sampling of the IMF. Furthermore, the (sometimes extreme) low-metallicity conditions in low-mass galaxies may affect dust production and evolution in ways that are not captured well by the dust models built for and largely calibrated on more massive galaxies.
Many other assumptions are involved in SED modeling (e.g., the stellar model library used, the fitting method), but we defer discussion of these to a later work.

We will address each of the four assumptions listed above in roughly \emph{independent} tests---for example, in Section~\ref{sec:sfhs} we will first fix the dust model and IMF assumptions to match the input SEDs, and will only vary the SFH model. 
In Section~\ref{sec:dust} we will fix the SFH model and IMF but vary the dust model; in Section~\ref{sec:imf} we will fix the SFH and dust models but vary the IMF.
This means that several of these tests are unrealistic---when performed on real observations, there is no way to fix the dust assumptions in Prospector to match a ``true'' dust model.
Perhaps the most realistic test of Prospector is in Section~\ref{sec:bands}, in which we assume the true dust attenuation law is unknown and make reasonable assumptions about the SFH template and IMF, then test how photometric coverage affects mass estimates.
We discuss potential systematic effects that may arise from the experimental setup of these tests in Section~\ref{sec:expsetup}.

The full priors used in all tests in this section are described in Appendix~\ref{sec:priors}.
Table~\ref{tab:residuals} summarizes the results of all tests in this section, including the mean and standard deviation of the residuals $\log(M_{\star,\mathrm{calib}}/M_{\star,\mathrm{true}})$ for low-mass galaxies from each model, as well as for all models combined.

As with any fitting procedure, the best-fit model identified by Prospector may not actually fit the data well. 
To test the goodness-of-fit, we use a version of the reduced chi-squared statistic $\chi_{\mathrm{red}}^2$ to compare the true synthetic SED with the SED prediction for the highest probability sample.\footnote{We define our reduced chi-squared statistic similar to the usual way, with $\chi_{\mathrm{red}}^2=\chi^{2}/\nu$; however, instead of $\nu$ representing the number of degrees of freedom $\nu=N_\mathrm{obs}-N_\mathrm{params}$, we set it to $\nu=N_\mathrm{obs}-1$ where $N_\mathrm{obs}$ is the number of observed photometric bands used in the fit. While in principle, the number of parameters in our SED fits is $>1$ (and is in fact $>N_\mathrm{obs}$), the goal of this study is to measure stellar mass, which is relatively robust to degeneracy (see discussion in Section~\ref{sec:hyperparams}).}
We identify fits with $\chi_{\mathrm{red}}^2>10$ as ``poor'' fits, marked with x's in all plots in this section.\footnote{Increasing (decreasing) the $\chi_{\mathrm{red}}^2$ cut-off for ``poor'' fits slightly decreases (increases) the fraction of ``poor'' fits for all tests, but does not affect our conclusions from this section.}
Table~\ref{tab:residuals} also lists the relative fraction of poor fits for each test, since our goal is to identify which prior assumptions most successfully recover $M_{\star}$; however, the ``poor'' fits are not considered in any further analysis (including the other summary statistics listed in Table~\ref{tab:residuals}).

\begin{figure*}[t!]
    \centering
    \epsscale{1.17}
    \plottwo{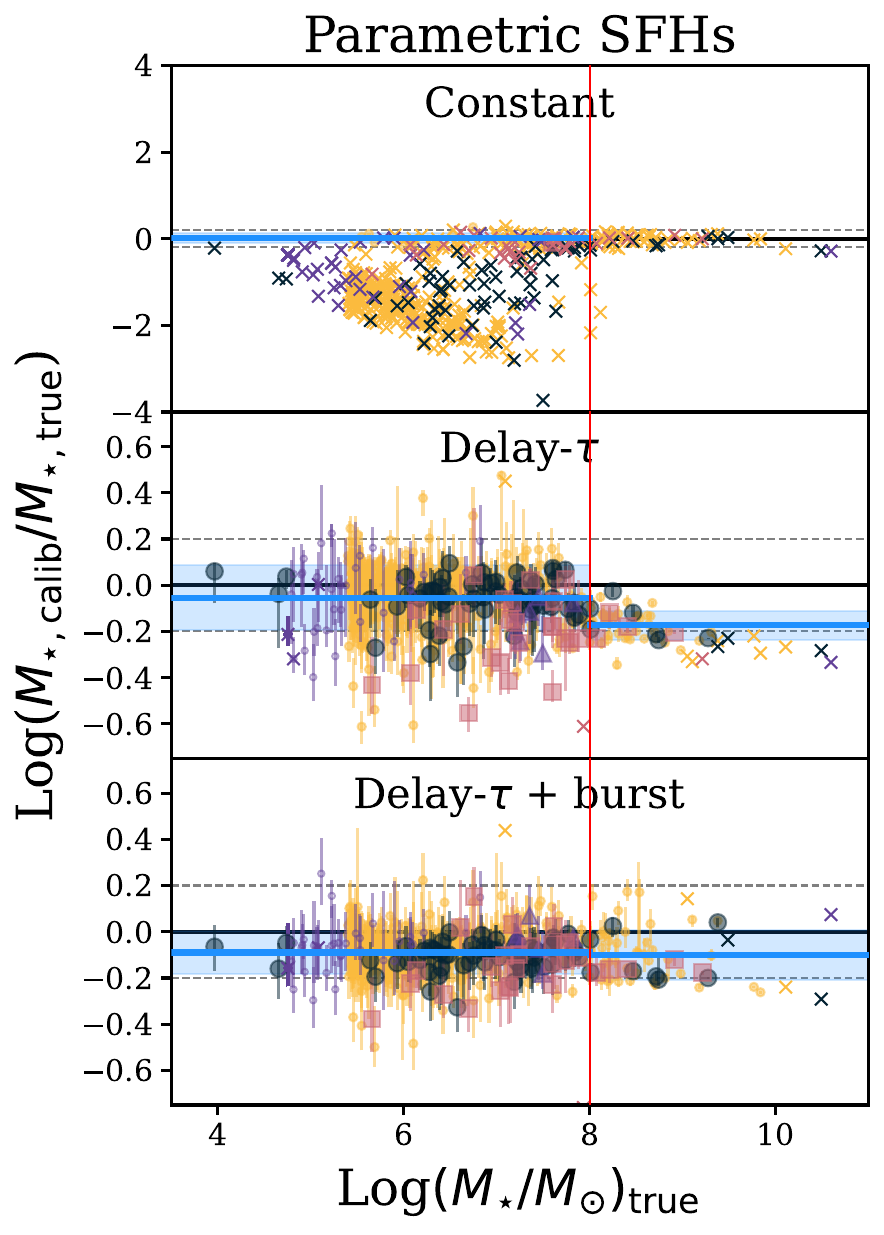}{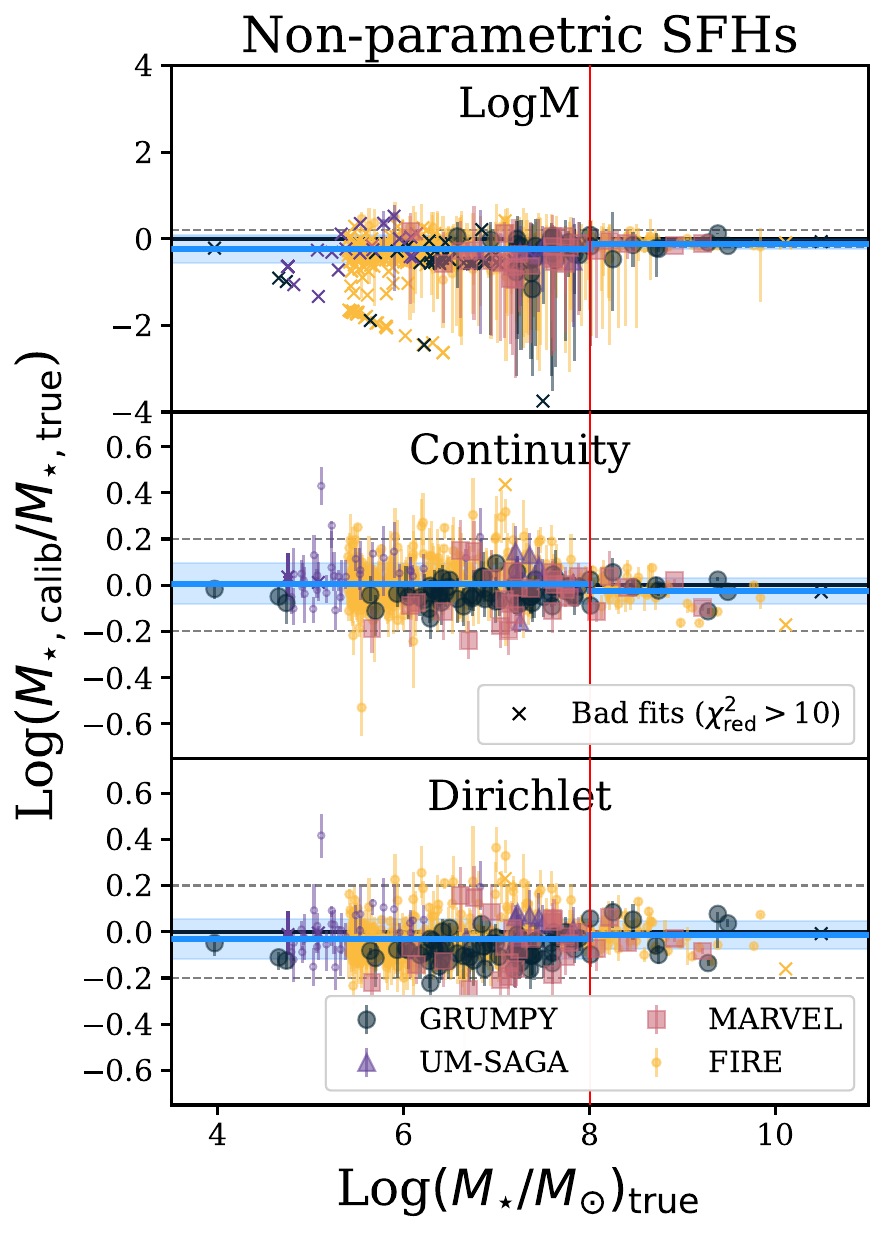}
    \caption{Same as Figure~\ref{fig:sim_empirical}, but $M_{\star,\mathrm{calib}}$ represents stellar masses measured using Prospector with various SFH priors. The left column shows the results of different parametric SFH priors: a constant SFH (top), a delayed-$\tau$ ($\sim t\exp^{-t/\tau}$) SFH (middle), and a delayed-$\tau$ SFH with a $\delta$-burst added (bottom). The right column shows the results of non-parametric SFH priors: a Log$M$ SFH (top), a continuity SFH (middle), and a Dirichlet SFH with a Dirichlet hyperparameter $\alpha=0.7$ (bottom). Bad fits ($\chi_{\mathrm{red}}^{2}>10$) are marked with x's. Note the expanded $y$-axis limits on the top plots.}
    \label{fig:sim_sfhs}
\end{figure*}

\subsection{Parametric vs. nonparametric SFHs}
\label{sec:sfhs}
We first test the effect of different assumptions about the shape of the SFH on stellar mass measurements.
We modify some of the default values in Prospector to accommodate the lower masses and metallicities expected for low-mass galaxies (see Appendix~\ref{sec:priors}).
For this test, we fix the dust attenuation, dust emission, and IMF to be the same as the input SEDs---we will relax these strict assumptions in the following sections.
We also use all 7 photometric bands listed in Table~\ref{tab:bands}, which tests the ``best-case'' scenario for low-$z$ galaxies in the SAGA and DESI surveys; we will investigate some of the effects caused by missing photometric coverage later in Section~\ref{sec:bands}.

The left column of Figure~\ref{fig:sim_sfhs} presents a comparison of Prospector's parametric SFH templates: constant SFH, delayed-$\tau$ SFH, and delayed-$\tau$ with an additional instantaneous burst of star formation.
We find that the assumption of a constant SFH (top left panel of Figure~\ref{fig:sim_sfhs}) leads to significant offsets for galaxies with $M_{\star}<10^{8}~\rm{M_{\odot}}$: using a constant SFH model, Prospector can underestimate true stellar masses by up to $\sim4$~dex.
This is not unexpected; \cite{Lower2020} also showed that a constant SFH can lead to similarly large offsets in $M_{\star}$ for more massive galaxies.
Additionally, we find that a constant SFH produces mostly ``poor'' fits: $\sim97\%$ of the fits produce $\chi_{\mathrm{red}}^{2}>10$.
Again, this is not surprising, since few of the $z\sim0$ galaxies in our sample have SFHs that are close to constant (see Fig.~\ref{fig:sim_sfhs}).
A constant SFH is more appropriate for high-redshift star-forming galaxies that are young enough to have maintained the same SFR over their short lifetimes.

For low-$z$ galaxies whose SFR has likely declined over time, a more realistic parametric SFH model is the delayed-$\tau$ model (left middle panel of Figure~\ref{fig:sim_sfhs}), in which SFR is described as a delayed declining exponential:
\begin{equation}
\mathrm{SFR}(t) \propto
\begin{cases}
     0 & t < T_{0} \\
    (t-T_{0})e^{-\frac{t-T_{0}}{\tau}} & t \geq T_{0}
\end{cases}
\label{eq:delaytau}
\end{equation}
where $T_{0}$ is the delay time of the SFR (or the maximum stellar age of the galaxy) and $\tau$ is the $e$-folding time.
For more massive galaxies ($M_{\star}>10^{8}~\rm{M_{\odot}}$), the delayed-$\tau$ model underpredicts the true stellar mass by $\sim0.2$~dex, in agreement with \cite{Lower2020}.
However, for lower-mass galaxies the systematic bias is much less dramatic; on average, after removing poor fits (which constitute only 3\% of our sample), stellar masses are only underpredicted by $-0.05$~dex on average, with a dispersion of $0.14$~dex.
As shown in the bottom left panel of Figure~\ref{fig:sim_sfhs}, increasing the flexibility of the model by adding an instantaneous $\delta$-function burst of star formation (characterized by the time of the burst and the fraction of stellar mass produced in the burst) further improves the recovery of $M_{\star}$ for low-mass galaxies (the percentage of poor fits drops even further, to 1.7\%) and reduces the dispersion in stellar mass residuals to $0.09$~dex.
Neither model shows a strong trend in residuals as a function of stellar mass below $M_{\star}<10^{8}~\rm{M_{\odot}}$.

We now consider ``nonparametric'' SFH models, which are described not with analytic functional forms but with hyperparameters---e.g., some number of time bins (which are each assumed to have constant SFR), the bin widths, and other hyperparameters describing the relationships between the SFRs in the bins.
The right column of Figure~\ref{fig:sim_sfhs} shows the performance of three of the SFH models implemented in Prospector \citep{Leja2017}.
For this test, we fix the number of bins to $N_{\mathrm{bin}}=10$ for all SFH models (see Section~\ref{sec:systematics} for further discussion of this choice).
Following the procedure of \citet{Lower2020}, the bins are logarithmically spaced except for the two youngest bins, which in this work are fixed to span lookback times of $0-10$~Myr and $10-100$~Myr.

The Log$M$ model is the simplest nonparametric model, in which the free parameters are the masses formed in each of the fixed time bins.
This SFH model is extremely flexible and can reproduce multiple bursts of star formation and quenching.  
However, as shown in the top right panel of Figure~\ref{fig:sim_sfhs}, the Log$M$ model struggles to accurately recover stellar masses. 
Only 39.5\% of the fits are ``good'' ($\chi_{\mathrm{red}}^2<10$); of these ``good'' fits, stellar mass is underestimated by $-0.24$~dex on average, with a dispersion of $0.32$~dex.

Nonparametric SFH templates with more constrained functional forms perform better than the Log$M$ model.
The Continuity model (right middle panel of Figure~\ref{fig:sim_sfhs}) fits directly for the difference in SFR between adjacent time bins, but explicitly weights against sharp transitions in SFR.
Although low-mass galaxy SFHs are thought to be highly stochastic and bursty, the Continuity model performs remarkably well, with an average stellar mass residual of $0.01$~dex and a dispersion of $0.09$~dex.
The Dirichlet model (bottom panel), like the Continuity model, also constrains the relationship between SFRs in each of the bins. 
In this model, the fractional sSFR for each time bin follows a Dirichlet distribution \citep{Betancourt2012}, set by a ``concentration'' parameter $\alpha$: low $\alpha$ values weight toward bursty SFHs, while higher $\alpha$ weights toward smooth SFHs.
We assume $\alpha=0.7$ following \citet{Lower2020}, although we discuss the effect of changing $\alpha$ in the following section.
The Dirichlet template is comparable to the Continuity model in terms of stellar mass recovery, with an average offset of $-0.03$~dex and a dispersion of $0.11$~dex.
Both the Continuity and the Dirichlet models return the highest number of ``good'' fits of all the SFH models (with only 1.28\% of all fits being discarded as ``poor'' fits), and neither template produces a strong trend in the residuals as a function of stellar mass.

The overall success of the Continuity and Dirichlet SFH models suggests that the additional flexibility provided by non-parametric SFHs improves stellar mass measurements of low-mass galaxies.
However, too much flexibility (as in the Log$M$ model) may under-constrain the fit, leading to poor SED fits and unreliable $M_\star$ measurements.
For the remainder of this work, we use the Dirichlet template as our fiducial SFH model; although the Continuity template performed similarly well, the Dirichlet model provides a more direct comparison to the work by \citet{Lower2020}, as discussed in the next section.

\subsubsection{Comparison with literature results}

These results are a direct extension of the work on higher-mass galaxies by \citet{Lower2020}, who demonstrated that for galaxies with $M_{\star}\gtrsim10^{8}~\rm{M_{\odot}}$, assuming simple parametric SFHs can lead to systematic biases in measured stellar mass of up to $\sim0.4$~dex.
Like this work, \citet{Lower2020} also used FSPS to generate mock photometry of simulated galaxies, then compared the ``true'' masses with Prospector's estimates.

There are some differences between the studies---most notably, \citet{Lower2020} used galaxies from a single large-volume hydrodynamic simulation \citep[SIMBA;][]{Dave2019}, while we used several different models that all specifically aim to model lower-mass galaxies (including zoom-in simulations that have much finer mass resolution than SIMBA).
Rather than using FSPS to apply a simple dust attenuation law while generating mock SEDs, as we have done here, \citet{Lower2020} used the more complex 3D radiative transfer code POWDERDAY to model a dust screen surrounding all stars.
Finally, while \citet{Lower2020} considered the effect of applying 3\% uncertainties to their mock photometry, we used realistic observational uncertainties that vary as a function of wavelength and for different galaxies.

Despite these slight differences in methods, many of our results are qualitatively in agreement.
We find that parametric SFHs generally perform worse than non-parametric SFHs, often underestimating the true stellar mass.
This is particularly apparent for the Constant SFH template: we find an large stellar mass offset of $-1.09\pm0.80$~dex for low-mass galaxies (when poor fits are included), which is consistent with the offset of $-0.48\pm0.61$~dex reported by \citet{Lower2020}. 
\citet{Lower2020} also find that other parametric SFHs (Delayed-$\tau$, Delayed-$\tau$+burst) tend to underestimate $M_\star$, which is qualitatively consistent with the high-mass ($M_\star>10^8~\rm{M_\odot}$) galaxies in our sample (right of the red vertical lines in the left middle and left bottom panels of Figure~\ref{fig:sim_sfhs}).

However, we find that for low-mass galaxies, parametric SFHs can still perform quite well: the Delayed-$\tau$ and Delayed-$\tau$+burst models produce stellar mass residuals that are, on average, consistent with zero within their 1~$\sigma$ dispersions. 
Additionally, \citet{Lower2020} find that the non-parametric Continuity SFH template overestimates stellar masses by $0.24\pm0.15$~dex, while we find that the Continuity template recovers masses that are much closer to the true mass ($0.01\pm0.09$~dex for the low-mass galaxies).
This is somewhat surprising: the parametric and Continuity templates prefer smoothly changing SFHs, but low-mass galaxies have burstier SFHs than massive galaxies (compare, e.g., the top panels of Figure~\ref{fig:samplesfh} with the bottom panel).

The exact reason for this discrepancy is unclear.
One option is that short-duration bursts of star formation simply do not contribute significantly to the overall stellar mass of a low-mass galaxy.
As a result, parametric and smooth SFHs may still recover a reasonably accurate stellar mass even if Prospector's SFH models are unable to capture the full behavior of the SFH.
We defer a full test of this hypothesis to future work.
In the meantime, we can at least confirm, as \citet{Lower2020} did, that non-parametric SFH models with some constraints (e.g., the Dirichlet and Continuity models) are the best among the Prospector SFH templates tested for stellar mass recovery.
Finally, because the Dirichlet and Continuity models performed comparably for our sample, for convenience we choose the Dirichlet model as our fiducial SFH model, since \citet{Lower2020} also adopted the Dirichlet model as their default non-parametric SFH.

\subsubsection{Non-parametric SFH hyperparameters}
\label{sec:hyperparams}
As described above, the non-parametric SFH models defined by Prospector are not truly ``non-parametric,'' in that they must still be described by hyperparameters such as the number and width of time bins ($N_{\mathrm{bins}}$) of constant SFH.
We therefore test the effect of changing $N_{\mathrm{bins}}$ for our fiducial SFH model (Dirichlet SFH with $\alpha=0.7$).
Following \citet{Ocvirk2006}, we use age bins spaced logarithmically in time. Notably, \citet{Ocvirk2006} recommend $N_{\mathrm{bins}}\leq8$ when measuring SFHs, since they find that $>8$ characteristic star-forming episodes cannot be recovered with high-quality optical spectra; however, since our goal is to measure stellar masses, which are far more robust than SFHs, some degeneracy from setting larger $N_\mathrm{bins}$ is acceptable.
Indeed, we find that if $N_\mathrm{bins}$ is too small, many of the fits are poor quality ($\chi^2_{\mathrm{red}}>10$); for example, with $N_\mathrm{bins}=3$, $70\%$ of the galaxies have poor fits. 
However, the ``good'' fits are still able to recover stellar masses that are on average consistent with the true stellar masses.
The number of poor fits can be reduced by increasing the number of bins, and the results for $N_\mathrm{bins}={6,10,12}$ are essentially the same.

We also test the effect of varying the Dirichlet $\alpha$ parameter, which determines the burstiness of the SFH.
Again, we find that if $\alpha$ is too small, the number of poor fits increases --- if $\alpha=0.2$, $41.8\%$ of the galaxies have poor fits --- but the ``good'' fits are still strong predictors of the true stellar mass. 
Increasing $\alpha$ reduces the number of poor fits, and $\alpha=0.7$ (the fiducial value) and $\alpha=1.0$ again return near-identical results.

These tests suggest that there are some minimum values for $N_\mathrm{bins}$ and $\alpha$, below which Prospector begins to struggle to fit many low-mass galaxy SFHs.
While determining these exact values is beyond the scope of this work (and is likely to depend on the sample of galaxies used --- or in this case, the details of the simulations used to model our galaxies), we can at least safely assume that the fiducial hyperparameters discussed in the previous section $N_\mathrm{bins}=10$ and $\alpha=0.7$ are reasonable assumptions.

\begin{figure*}[t!]
    \centering
    \epsscale{1.17}
    \plottwo{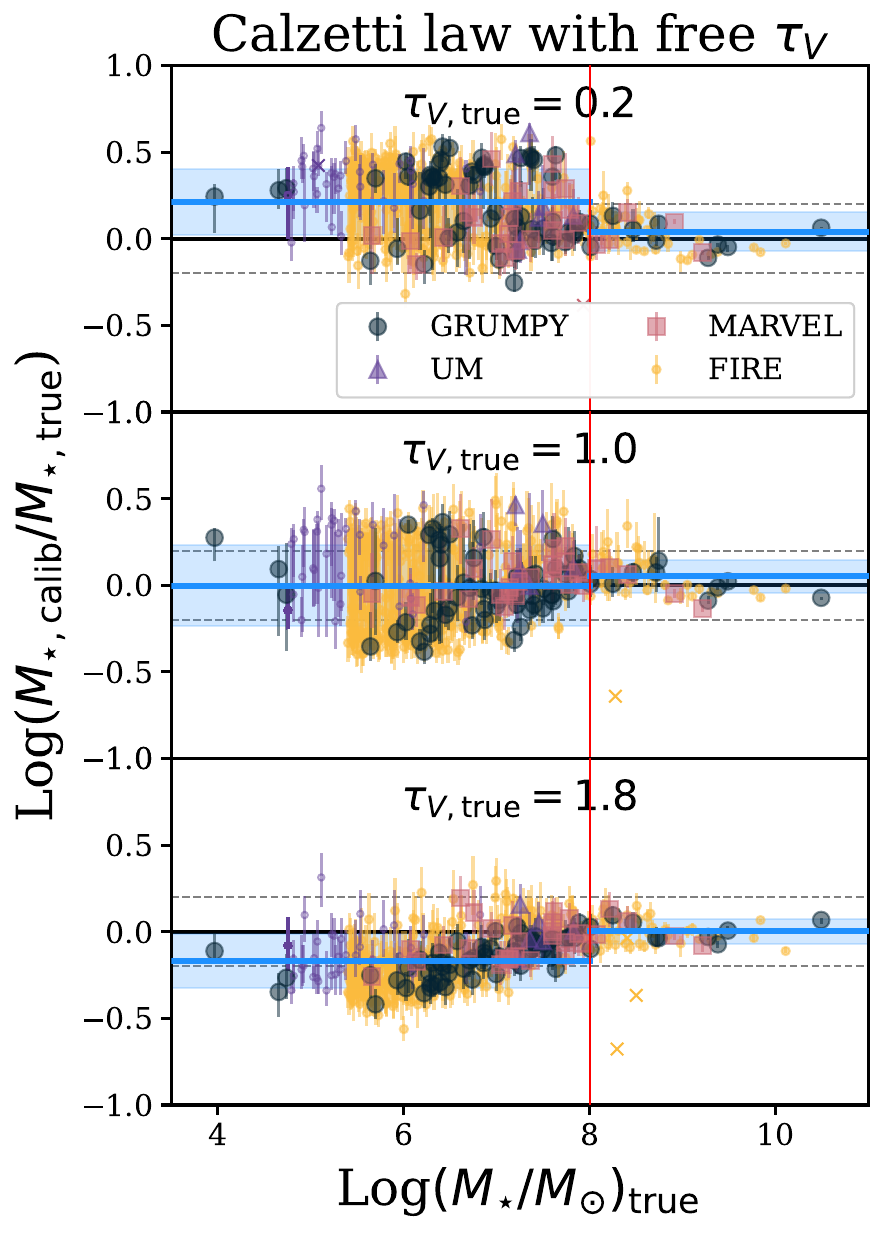}{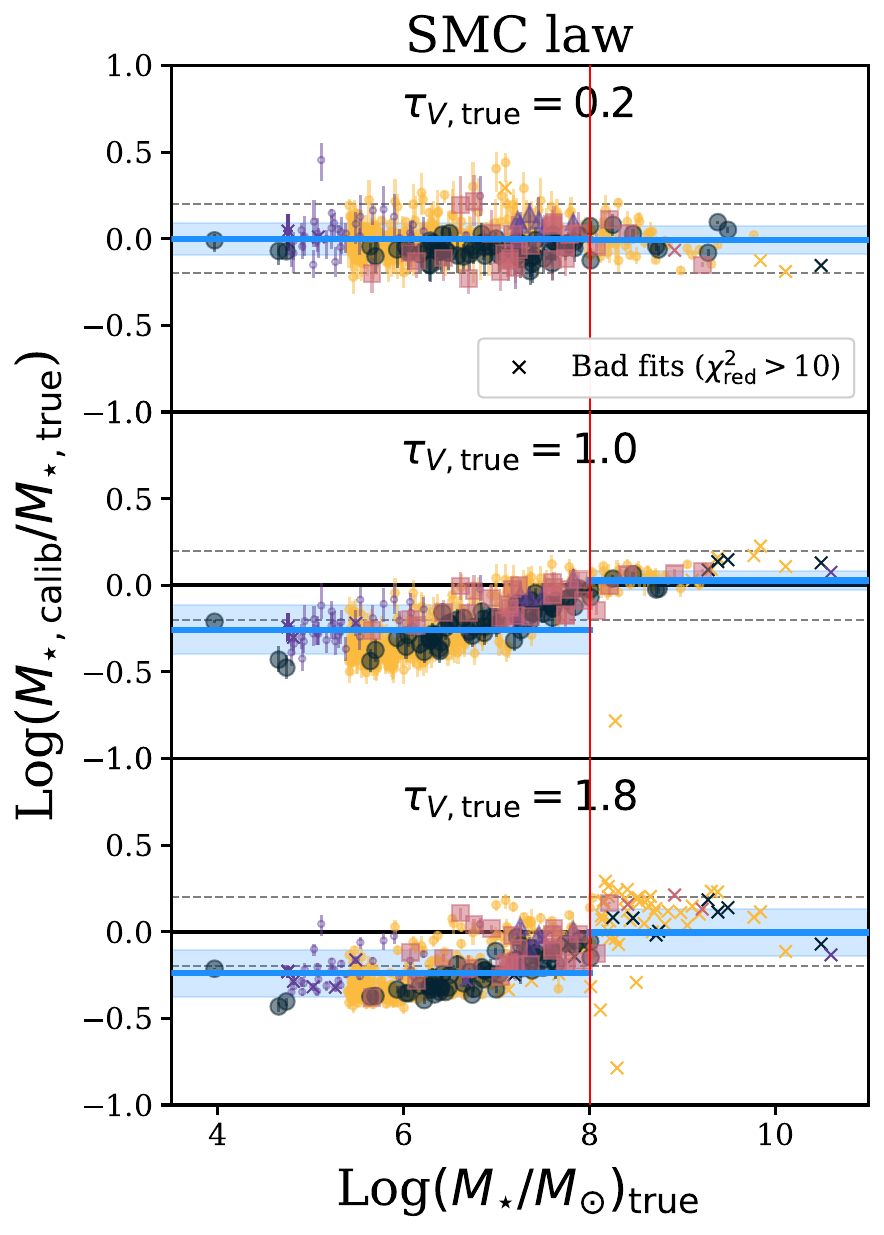}
    \caption{Same as Figure~\ref{fig:sim_empirical}, but $M_{\star,\mathrm{calib}}$ represents stellar masses measured using Prospector with different dust attenuation laws. The rows represent mock observations with varying amounts of dust, produced by using the \citet{Calzetti2000} dust law with different normalization parameters $\tau_{V,\mathrm{true}}$: 0.2 (top), 1.0 (middle), and 1.8 (bottom).
    The left column shows the results when assuming a \citet{Calzetti2000} dust law with the normalization parameter $\tau_{V}$ allowed to vary as a free parameter. The right column shows the results when assuming a SMC bar attenuation law \citep{Gordon2003}. Bad fits ($\chi_{\mathrm{red}}^{2}>10$) are marked with x's.}
    \label{fig:sim_dustatt}
\end{figure*}

\subsection{Dust prescription}
\label{sec:dust}
In the previous sections, we tested SFH assumptions by using idealized models in which the dust attenuation and emission exactly matched the input SEDs.
Of course, this is not the case in reality.
Having identified a fiducial SFH model, we can now address the impact of dust on our SED fits.

Dust both attenuates and emits light from a galaxy, generally leading to a reddening of galaxy spectra.
While significant work has been done to progress our understanding of both of these effects --- see, e.g., \citet{Salim2020} for a recent review of the dust attenuation law --- the impact of dust on galaxy observations remains an open question, particularly in low-mass and low-metallicity galaxies.
This is both because the physical properties of dust in low-metallicity environments are not well understood \citep[e.g.,][]{Nanni2020,Galliano2021}, and because dust, metallicity, and age have inherently degenerate effects on a galaxy's SED \citep[and this degeneracy is also a complex function of galaxy mass and metallicity; e.g.,][]{Nagaraj2022}.

In this work, we do not attempt to choose the most physically correct global dust prescription for low-mass galaxies.
Our goal is simply to test the robustness of Prospector's ability to recover $M_\star$ of low-mass galaxies, given different assumptions about dust models.
To do this, we vary the dust model used to produce the mock observations, then run Prospector with different parameterizations of the dust attenuation model.
For these tests, we fix the SFH to our fiducial nonparametric Dirichlet SFH with $\alpha=0.7$.

\subsubsection{Dust attenuation}

We first test the impact of the assumed dust attenuation model.
We consider both (1) the overall normalization of the attenuation law, and (2) the shape of the attenuation law.
Again, we test these independently: we first fix the shape of the dust attenuation law and vary the normalization, then we fix the normalization and vary the shape.

As described in Section~\ref{sec:mockobs}, our primary set of mock observations are produced by assuming a \citet{Calzetti2000} attenuation law normalized to $\tau_{V}=0.2$.
To test how strongly the \emph{normalization} of the dust attenuation curve affects stellar mass measurements, we synthesize additional sets of observations using a \citet{Calzetti2000} law normalized to $\tau_{V}=1.0$ and $1.8$.
We then fit these mock SEDs using the \citet{Calzetti2000} law, but where $\tau_V$ is allowed to vary as a free parameter.
This parameterization of the dust attenuation law is a common assumption when fitting the SEDs of low-mass galaxies \citep[e.g.,][]{Pandya2018,Greco2018}, although the exact form of the prior on $\tau_V$ may vary; in this case, we assume a uniform prior for $0<\tau_V<$2. 

The results of changing the dust normalization are shown in the left column of Figure~\ref{fig:sim_dustatt}, where the true value of $\tau_{V}$ (i.e., the input value used to produce the mock observations) increases from the top row to the bottom row.
Two primary features stand out: first, regardless of the true $\tau_{V}$ of the mock observations, there is a increase in scatter in the recovered stellar mass residuals below $M_{\star,\mathrm{true}}<10^8~\rm{M_{\odot}}$.
This is visibly apparent in the left column of Figure~\ref{fig:sim_dustatt}; the blue horizontal bands to the left of the red vertical lines are larger than the blue horizontal bands to the right of the vertical lines.
This increase in scatter is the direct result of allowing the dust normalization to vary: for low-mass galaxies, the dispersion in stellar mass increases from $0.09$~dex when the dust prescription exactly matches the input dust law (bottom right panel of Figure~\ref{fig:sim_sfhs}) to $[0.19,0.23,0.15]$~dex when $\tau_{V}$ is allowed to vary (for $\tau_{V,\mathrm{true}}=[0.2,1.0,1.8]$, respectively).

\begin{figure}[t!]
    \centering
    \epsscale{1.16}
    \plotone{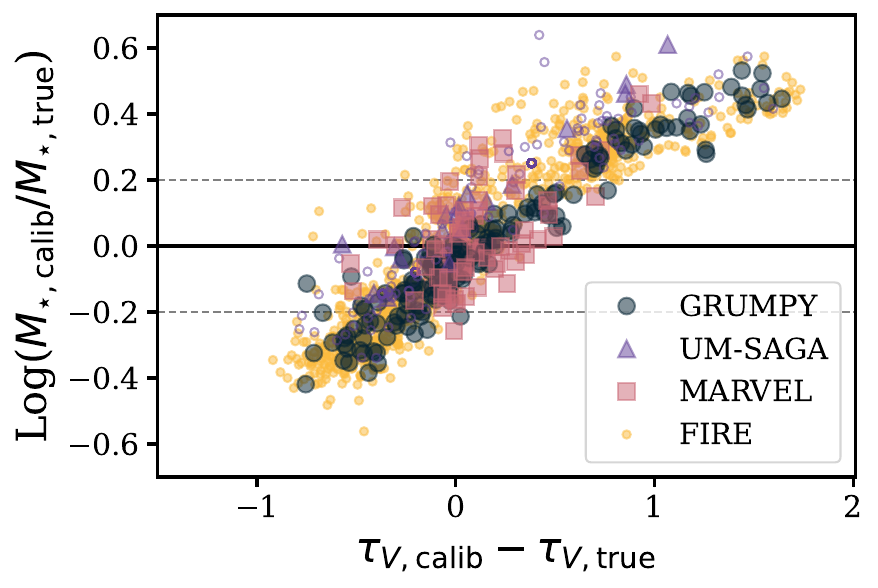}
    \caption{Difference between measured and actual stellar mass $\log(M_{\star,\mathrm{calib}}/M_{\star,\mathrm{true}})$ as a function of the difference between measured dust normalization $\tau_{V,\mathrm{calib}}$ and actual dust normalization $\tau_{V,\mathrm{true}}$. The dashed horizontal gray lines mark $\pm0.2$~dex uncertainties in stellar mass. UM-SAGA galaxies with $\log(M_\star/\rm{M_\odot})<7$ are marked with small unfilled points to denote that their SFHs are poorly constrained.}
    \label{fig:dustcheck}
\end{figure}

Additionally, the average residuals for low-mass galaxies (the blue horizontal bands left of the red vertical lines in Figure~\ref{fig:sim_dustatt}) appear to be a function of $\tau_{V,\mathrm{true}}$.
The free-$\tau_{V}$ models typically overestimate stellar masses (mean offset of $0.21$~dex) when $\tau_{V,\mathrm{true}}=0.2$ and underestimate stellar masses (mean offset of $-0.17$~dex) when $\tau_{V,\mathrm{true}}=1.8$.
This makes sense, since we stipulated in our prior assumptions that $0<\tau_{V}<2$.
As a result, if $\tau_{V,\mathrm{true}}=0.2$ ($\tau_{V,\mathrm{true}}=1.8$), Prospector is more likely to overestimate (underestimate) $\tau_{V}$, and overestimating the total amount of dust will lead to an overestimate (underestimate) of the true stellar mass.
To check this hypothesis, Figure~\ref{fig:dustcheck} plots the residuals in stellar mass as a function of the residuals in $\tau_{V}$ (i.e., the difference between the measured and ``true'' $\tau_{V}$ values). 
As expected, there is a clear trend between the two residuals: overestimating $\tau_{V}$ ($\tau_{V,\mathrm{calib}}-\tau_{V,\mathrm{true}}>0$) is correlated with overestimating $M_{\star}$, and underestimating $\tau_{V}$ is correlated with underestimating $M_{\star}$.

We can also test the effect of changing the \emph{shape} of the dust attenuation law by fitting our mock SEDs with the extinction law measured by \citet{Gordon2003} for the SMC bar, which has a much steeper UV-optical slope than the \citet{Calzetti2000} law.
Rather than a single normalization parameter, as was used with the \citet{Calzetti2000} law, we now assume a dust screen geometry in which young stars are more strongly attenuated by their birth clouds, while older stellar populations are less strongly attenuated by diffuse cirrus dust.
Prospector uses the \citet{CharlotFall2000} parameterization to model how opacity varies as a function of stellar age $t$.
We assume a power-law dust screen for young stars (with $t \leq10^{7}$~y), while both young and old stars are attenuated by diffuse dust:
\begin{align}
    \tau_{\mathrm{young}} & = \tau_1(\lambda/5500\mathrm{\AA})^{-\delta_\mathrm{CF}} \\
    \tau_{\mathrm{diff}} & = \tau_2(A(\lambda)/A(V))
\end{align}
Following \citet{Conroy2010b} and \citet{CharlotFall2000}, we set $\delta_\mathrm{CF}=1.0$, $\tau_1=1.0$, and $\tau_2=0.3$.
We use fixed values of $\tau_1$ and $\tau_2$ because our goal is simply to test the effect of changing the shape of the dust attenuation law, not the normalization.
The diffuse dust attenuation $A(\lambda)/A(V)$ is set by the SMC bar extinction law \citep[Table 4 of][]{Gordon2003}.

The results of changing the shape of the dust attenuation law are shown in the right column of Figure~\ref{fig:sim_dustatt}, where the SMC dust law is applied to mock SEDs produced using the \citet{Calzetti2000} law with varying $\tau_{V,\mathrm{true}}$.
As before, $\tau_{V,\mathrm{true}}$ traces the overall amount of dust.
For galaxies with low amounts of dust ($\tau_{V,\mathrm{true}}=0.2$; top row), the exact form of the dust attenuation law does not significantly change the output; assuming the SMC dust law produces stellar masses that are as accurate as if the ``true'' dust law was used ($0.00\pm0.09$~dex, compared to $-0.03\pm0.09$~dex when the dust law is fixed to match the input).
However, as the total amount of dust increases, using the incorrect dust law more dramatically impacts the recovered stellar masses.
For both $\tau_{V,\mathrm{true}}=1.0$ and $\tau_{V,\mathrm{true}}=1.8$, the SMC dust law underpredicts the stellar masses of $M_\star<10^8~\rm{M_\odot}$ galaxies by $\sim-0.3$~dex on average. 
Furthermore, as the true value of $\tau_{V}$ increases, the number of ``poor'' fits increases, particularly for the most massive galaxies in our sample.

\subsubsection{Dust emission}
We also check the effect of the dust emission model on stellar mass measurements. 
We again fix the SFH to a Dirichlet prior, this time with the dust attenuation law fixed to be exactly the same as the input SEDs \citep[][with $\tau_{V}=0.2$]{Calzetti2000}. 
In this test, we allow the dust emission parameters in the \citet{Draine2007} dust model --- the minimum radiation field $U_{\mathrm{min}}$, the fraction of dust heated by starlight $\gamma$, and the PAH mass fraction $q_{\mathrm{PAH}}$ --- to vary.
Compared to the same model in which these parameters were fixed (right bottom panel of Figure~\ref{fig:sim_sfhs}), this additional flexibility improves the fits for some galaxies, particularly those with stellar masses $\gtrsim10^{8}~\rm{M_{\odot}}$; the percentage of ``poor'' fits decreases to 0.64\%.
However, for low-mass galaxies with $M_{\star,\mathrm{true}}<10^{8}~\rm{M_{\odot}}$, varying the dust emission parameters only slightly changes the recovered stellar masses, and the overall mean and standard deviation of the residuals do not change at all.
This is almost certainly because we have limited our SEDs to UV through near-IR (and the reddest band we use, W2, is particularly noisy; see Section~\ref{sec:litcalibs}), while dust emission peaks in the mid- and far-IR.
Dust emission assumptions likely have a more significant effect on stellar mass measurements based on SEDs with redder coverage.

\vspace{1em}
To summarize, the assumed dust model has a strong impact on both the systematic offset and the scatter in stellar masses recovered from SED fitting, and these effects are particularly apparent for low-mass ($M_{\star}<10^8~M_{\odot}$) galaxies:
\begin{itemize}
    \item Assumptions about the overall \emph{normalization} of the attenuation law (i.e., the total amount of dust) affect the recovered stellar mass in two ways. 
    \begin{enumerate}
        \item Allowing dust normalization to vary increases the population-level scatter in recovered masses for low-mass galaxies.
        \item Any offset in the prior on dust normalization will lead to systematic offsets in the recovered stellar masses, because overestimating (underestimating) the dust normalization will lead to overestimating (underestimating) $M_\star$. Since most low-mass galaxies are likely to have low metallicities and correspondingly low dust masses, a wide prior on the dust normalization is more likely to \emph{overestimate} the amount of dust and consequently overestimate $M_\star$.
    \end{enumerate}
    \item The \emph{shape} of the dust attenuation law can also have an effect on the recovered stellar masses, but this only becomes a significant issue for dusty galaxies with $\tau_{V}>0.2$; again, since most low-mass galaxies may have low dust content, this may not be a major issue in this mass regime.
    \item Assumptions about dust \emph{emission} do not appear to significantly impact the results of SED fitting to UV through near-IR bands, although we note that it may be more relevant when mid- and far-IR photometry is included.
\end{itemize}

\subsection{Stellar initial mass function}
\label{sec:imf}
We briefly consider the role of the stellar IMF in SED fitting.
The most common IMFs assumed for nearby low-mass galaxies are those of \citet{Kroupa2001}, \citet{Chabrier2003}, and \citet{Salpeter1955}.
These primarily differ in shape at the low-mass end; since low-mass stars provide the dominant contribution to the stellar mass of a galaxy, but not its luminosity (compared to high-mass stars), changes to the low-mass IMF should, to first order, add a constant offset to $\log M_{\star}$.
We note that the Kroupa/Chabrier IMFs have slightly lower slopes at the high-mass end than the Salpeter IMF, leading to changes in luminosity and color, so the $M_\star/L$ of stellar populations using the Salpeter IMF may vary as a function of age and metallicity relative to Kroupa/Chabrier IMFs \citep[e.g., see Fig 4 of][]{Madau2014} --- however, these variations are much weaker ($\lesssim0.05$~dex) than the average offset in stellar mass.
On average, fitting models with a Kroupa IMF to mock observations produced using a Chabrier IMF should overestimate stellar masses by $\sim0.03$~dex, while fitting a Salpeter IMF to a Chabrier IMF should overestimate stellar masses by $\sim0.21$~dex.

We check this intuition by applying different IMFs in Prospector, assuming the fiducial Dirichlet SFH and dust parameters fixed to match the input SEDs.
As expected, we find that changing the IMF leads to near-constant shifts in stellar mass residuals (see Table~\ref{tab:residuals}): the average offset shifts, but the 1$\sigma$ dispersion does not change.
The magnitudes of these offsets are also smaller than expected from SSP models.
Compared to the true \citet{Chabrier2003} IMF, assuming a \citet{Kroupa2001} IMF
changes the average offset in stellar mass by only $0.01$~dex (increasing from $-0.03\pm0.09$~dex to $-0.02\pm0.09$~dex), while assuming a \citet{Salpeter1955} IMF increases the stellar mass residuals by $0.14$~dex on average (increasing from $-0.03\pm0.09$~dex to $0.11\pm0.09$~dex).

\subsection{Photometric coverage}
\label{sec:bands}
In the previous subsections, we considered only the results of fitting to all 7 photometric bands in Table~\ref{tab:bands}. 
In real low-$z$ galaxy surveys, incomplete photometric coverage is common. 
We now investigate how well Prospector can recover stellar masses with realistic subsets of the 7 photometric bands.
For this test, we run Prospector on our fiducial set of mock observations (produced using a \citealt{Chabrier2003} IMF and a \citealt{Calzetti2000} attenuation law with $\tau_{V}=0.2$). 
We aim to mimic a more ``realistic'' usage of Prospector, in which the true dust attenuation law is unknown and must be described by one or more free parameters.
However, some basic assumptions can be made: due to low-mass galaxies' low metallicities, one might expect these galaxies to have relatively little dust --- indeed, recent observations suggest that low-mass galaxies in the nearby universe have $0\lesssim\tau_V\lesssim0.4$ \citep[e.g.,][]{Geha2024, Greco2018, Pandya2018}.
As discussed in Section~\ref{sec:dust}, for galaxies with such low dust content, the exact form of the assumed dust attenuation law does not strongly impact the recovered stellar mass.
We therefore assume a \citet{Calzetti2000} attenuation law with a more restricted uniform prior on the normalization parameter $\tau_V$: $\mathcal{U}(0,0.4)$.

We then fit Prospector to three different sets of bands: (1) only optical $grz$ bands, (2) optical $grz$ with near-IR coverage (WISE W1 and W2), and (3) optical $grz$ with limited near-UV and near-IR coverage (GALEX NUV and WISE W1). 
The first case is motivated by the DESI Imaging Legacy Surveys \citep{Dey2019}, an extremely wide-field ($\approx14,000~\deg^{2}$) $grz$ survey.
The second case is motivated by the fact that while the WISE satellite imaged the entire sky in four near-to-mid-IR bands at 3.4, 4.6, 12, and 22~$\mu$m \citep[W1, W2, W3 and W4;][]{Wright2010,Cutri2012}, substantially deeper coverage is available for the two bluest bands, W1 and W2. 
Finally, the third case is motivated by the SAGA survey \citep{Mao2024,Geha2024}, which uses GALEX NUV fluxes to measure star formation rates.

\begin{figure}[t!]
    \centering
    \epsscale{1.18}
    \plotone{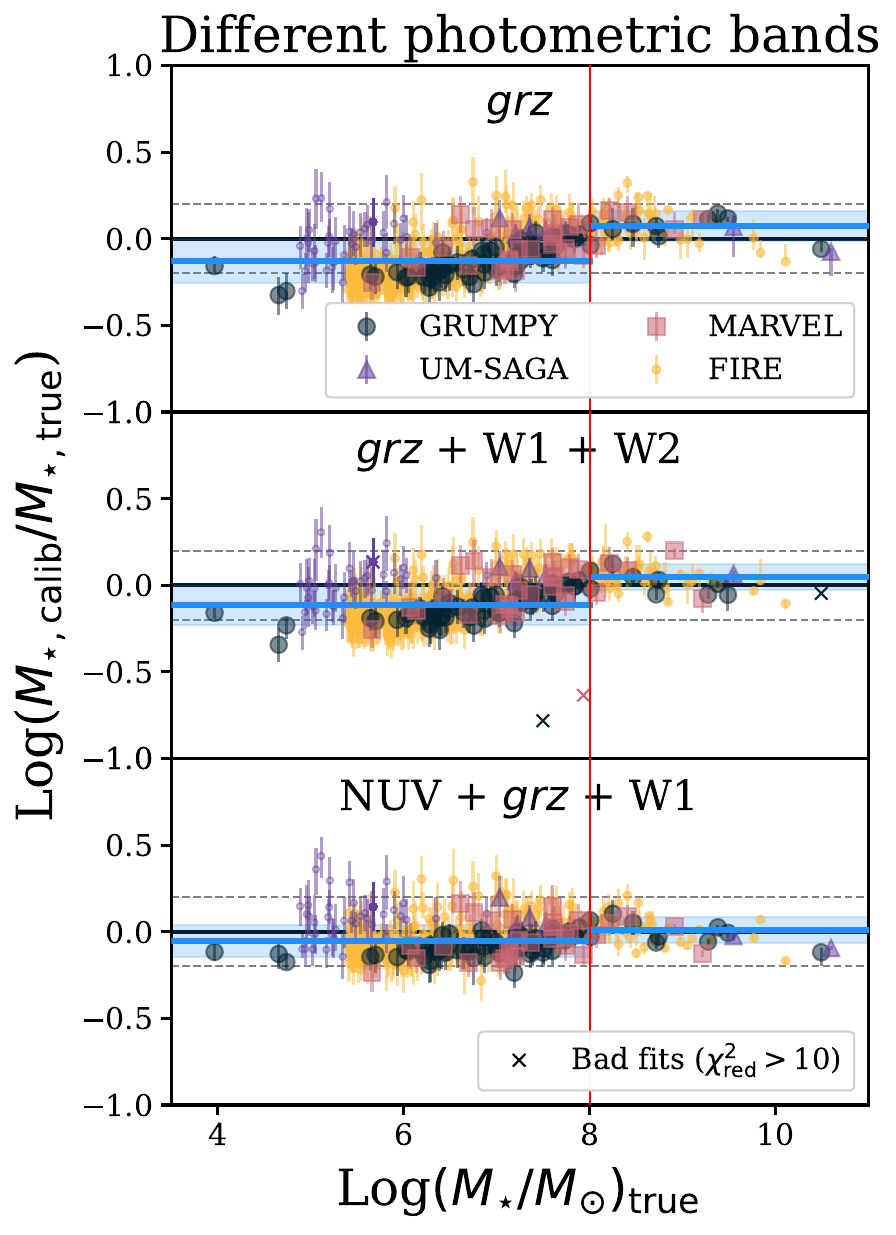}
    \caption{Same as Figure~\ref{fig:sim_empirical}, but $M_{\star,\mathrm{calib}}$ represents stellar masses measured using Prospector with different sets of photometric bands: optical $grz$ only (top), optical $grz$ and near-IR (middle), and optical $grz$ with limited near-UV and near-IR coverage (bottom).}
    \label{fig:sim_bands}
\end{figure}

The results of this test are shown in Figure~\ref{fig:sim_bands}.
As illustrated in the top panel, the optical $grz$ bands alone are able to recover much of the stellar mass information for low-mass galaxies.
While the $grz$ bands produce a systematic trend in stellar mass residuals as a function of $M_\star$, similar to the results of the empirical $g-r$ calibrations (Figure~\ref{fig:sim_empirical}), on average the low-mass residuals are consistent with zero ($-0.13\pm0.13$~dex).
Adding more bands slightly improves the $M_\star$ estimates for low-mass galaxies: the average residuals decrease to $-0.11\pm0.12$~dex if W1 and W2 are added, and to $-0.05\pm0.09$~dex if NUV and W1 are added.
The NUV+$grz$+W1 test shows no systematic trend in residuals as a function of stellar mass.
However, it is worth noting that the inclusion of the W2 band produces ``poor'' ($\chi^2_{\mathrm{red}}>10$) fits, likely due to the large W2 photometric uncertainties in our sample.
We recommend that future studies exercise caution in using uncertain photometric measurements when fitting low-mass galaxy SEDs.
Near-IR fluxes from JWST or other facilities may be more precise than WISE W2 fluxes and should provide more useful constraints on stellar masses.

\begin{deluxetable*}{lllllll}
\tablecolumns{7} 
\tablecaption{Residuals between ``true'' and measured masses, $\mathrm{log}\left(M_{\star,\mathrm{calib}}/M_{\star,\mathrm{true}}\right)$, for different SED fitting tests. The values reported are the mean and standard deviation of the residuals for low-mass ($M_{\star,\mathrm{true}} < 10^{8}~M_{\odot}$) galaxies. Only ``good'' fits ($\chi_{\mathrm{red}}^2<10$) are included when computing the mean and 1$\sigma$ residuals; the percentage of ``poor'' fits ($\chi_{\mathrm{red}}^2>10$) out of 645 total fits is also listed. \label{tab:residuals}} 
\tablehead{ 
\multicolumn{7}{c}{SED fitting with parametric SFHs} \\
\colhead{SFH prior} & 
\colhead{GRUMPY} & 
\colhead{UM-SAGA} & 
\colhead{MARVEL} &
\colhead{FIRE} & 
\colhead{All models} &
\colhead{\% bad}
}
\startdata
Constant SFH & $\ldots$ & $\ldots$ & $-0.08\pm0.00$ & $\nm0.02\pm0.11$ & $0.02\pm0.11$ & 97.4\% \\
Delayed-$\tau$ SFH & $-0.06\pm0.09$ & $-0.16\pm0.08$ & $-0.21\pm0.16$ & $-0.04\pm0.14$ & $-0.05\pm0.14$ & 2.99\% \\
Delayed-$\tau$+burst SFH & $-0.11\pm0.06$ & $-0.04\pm0.07$ & $-0.13\pm0.12$ & $-0.08\pm0.09$ & $-0.09\pm0.09$ & 1.71\% \vspace{0.5em}\\
\hline
\hline
\multicolumn{7}{c}{SED fitting with nonparametric SFHs} \\
\colhead{SFH prior} & 
\colhead{GRUMPY} & 
\colhead{UM-SAGA} & 
\colhead{MARVEL} &
\colhead{FIRE} & 
\colhead{All models} &
\colhead{\% bad} \\
\hline
Log$M$ SFH & $-0.29\pm0.31$ & $-0.34\pm0.11$ & $-0.38\pm0.30$ & $-0.18\pm0.31$ & $-0.24\pm0.32$ & 60.55\% \\
Continuity SFH & $-0.03\pm0.04$ & $\nm0.04\pm0.09$ & $-0.04\pm0.10$ & $\nm0.02\pm0.09$ & $\nm0.01\pm0.09$ & 1.28\% \\
\textbf{Dirichlet SFH}\tablenotemark{a} & $\mathbf{-0.08\pm0.05}$ & $\mathbf{\nm0.04\pm0.04}$ & $\mathbf{-0.06\pm0.11}$ & $\mathbf{-0.02\pm0.09}$ & $\mathbf{-0.03\pm0.09}$ & \textbf{1.28\%} \vspace{0.5em}\\
\hline
\hline
\multicolumn{7}{c}{SED fitting with different Dirichlet hyperparameters} \\
\colhead{SFH prior} & 
\colhead{GRUMPY} & 
\colhead{UM-SAGA} & 
\colhead{MARVEL} &
\colhead{FIRE} & 
\colhead{All models} &
\colhead{\% bad} \\
\hline
$N_\mathrm{bins}=3$ & $-0.09\pm0.06$ & $\nm0.06\pm0.03$ & $-0.05\pm0.12$ & $\nm0.02\pm0.12$ & $\nm0.00\pm0.12$ & 70.0\% \\
$N_\mathrm{bins}=6$ & $-0.10\pm0.06$ & $\nm0.04\pm0.08$ & $-0.06\pm0.11$ & $-0.07\pm0.11$ & $-0.07\pm0.10$ & 1.28\% \\
$N_\mathrm{bins}=12$ & $-0.09\pm0.05$ & $-0.02\pm0.09$ & $-0.06\pm0.11$ & $-0.03\pm0.09$ & $-0.04\pm0.09$ & 1.28\% \\
\hline
$\alpha=0.2$ & $-0.07\pm0.07$ & $-0.01\pm0.09$ & $-0.12\pm0.12$ & $-0.03\pm0.08$ & $-0.04\pm0.09$ & 41.8\% \\
$\alpha=1.0$ & $-0.09\pm0.05$ & $\nm0.01\pm0.10$ & $-0.06\pm0.12$ & $-0.03\pm0.09$ & $-0.04\pm0.09$ & 1.28\% \vspace{0.5em}\\
\hline
\hline
\multicolumn{7}{c}{SED fitting with different dust attenuation models} \\
\colhead{Dust models} & 
\colhead{GRUMPY} & 
\colhead{UM-SAGA} & 
\colhead{MARVEL} &
\colhead{FIRE} & 
\colhead{All models} &
\colhead{\% bad} \\
\hline
\multicolumn{7}{c}{\emph{Input SEDs with \citet{Calzetti2000} ($\tau_{V}=0.2$)}} \\
Calzetti, $\tau_V$ free & $\nm0.21\pm0.21$ & $\nm0.21\pm0.23$ & $\nm0.11\pm0.16$ & $\nm0.22\pm0.18$ & $\nm0.21\pm0.19$ & 0.85\% \\
SMC \citep{Gordon2003} & $-0.06\pm0.05$ & $\nm0.08\pm0.05$ & $-0.04\pm0.12$ & $\nm0.01\pm0.09$ & $\nm0.00\pm0.09$ & 1.71\% \\
\hline
\multicolumn{7}{c}{\emph{Input SEDs with \citet{Calzetti2000} ($\tau_{V}=1.0$)}} \\
Calzetti, $\tau_V$ free & $-0.02\pm0.19$ & $\nm0.18\pm0.17$ & $\nm0.05\pm0.12$ & $-0.01\pm0.25$ & $\nm0.00\pm0.23$ & 0.64\% \\
SMC \citep{Gordon2003} & $-0.22\pm0.11$ & $-0.08\pm0.07$ & $-0.11\pm0.08$ & $-0.28\pm0.14$ & $-0.26\pm0.14$ & 2.77\% \\
\hline
\multicolumn{7}{c}{\emph{Input SEDs with \citet{Calzetti2000} ($\tau_{V}=1.8$)}} \\
Calzetti, $\tau_V$ free & $-0.16\pm0.11$ & $-0.03\pm0.10$ & $-0.06\pm0.11$ & $-0.18\pm0.16$ & $-0.17\pm0.15$ & 0.85\% \\
SMC \citep{Gordon2003} & $-0.24\pm0.10$ & $-0.07\pm0.08$ & $-0.13\pm0.11$ & $-0.25\pm0.14$ & $-0.24\pm0.14$ & 13.43\% \vspace{0.5em}\\
\hline
\hline
\multicolumn{7}{c}{SED fitting with free dust emission parameters} \\
\colhead{} & 
\colhead{GRUMPY} & 
\colhead{UM-SAGA} & 
\colhead{MARVEL} &
\colhead{FIRE} & 
\colhead{All models} &
\colhead{\% bad} \\
\hline
Varying dust emission & $-0.08\pm0.05$ & $0.06\pm0.06$ & $-0.06\pm0.12$ & $-0.02\pm0.09$ & $-0.03\pm0.09$ & 0.64\% \vspace{0.5em}\\
\hline
\hline
\multicolumn{7}{c}{SED fitting with different IMFs} \\
\colhead{Model} & 
\colhead{GRUMPY} & 
\colhead{UM-SAGA} & 
\colhead{MARVEL} &
\colhead{FIRE} & 
\colhead{All models} &
\colhead{\% bad} \\
\hline
\citet{Kroupa2001} & $-0.07\pm0.05$ & $\nm0.05\pm0.05$ & $-0.06\pm0.10$ & $-0.01\pm0.09$ & $-0.02\pm0.09$ & 1.28\% \\
\citet{Salpeter1955} & $\nm0.06\pm0.04$ & $\nm0.19\pm0.04$ & $\nm0.10\pm0.11$ & $\nm0.12\pm0.09$ & $\nm0.11\pm0.09$ & 1.28\% \vspace{0.5em} \\
\hline
\hline
\multicolumn{7}{c}{SED fitting with different photometric bands} \\
\colhead{Model} & 
\colhead{GRUMPY} & 
\colhead{UM-SAGA} & 
\colhead{MARVEL} &
\colhead{FIRE} & 
\colhead{All models} &
\colhead{\% bad} \\
\hline
$grz$ only & $-0.13\pm0.09$ & $\nm0.06\pm0.04$ & $-0.05\pm0.10$ & $-0.14\pm0.13$ & $-0.13\pm0.13$ & 0\% \\
$grz$ + W1 + W2 & $-0.12\pm0.08$ & $\nm0.07\pm0.03$ & $-0.04\pm0.11$ & $-0.12\pm0.12$ & $-0.11\pm0.12$ & 1.07\% \\
NUV + $grz$ + W1 & $-0.09\pm0.05$ & $\nm0.08\pm0.07$ & $-0.05\pm0.11$ & $-0.05\pm0.10$ & $-0.05\pm0.09$ & 0\% \vspace{0.5em} \\
\enddata
\tablenotetext{a}{This model is the ``fiducial'' SED fit, in which the dust attenuation, dust emission, and IMF are all fixed to match the input SEDs. This model uses a Dirichlet SFH with $\alpha=0.7$ and $N_\mathrm{bins}=10$.}
\end{deluxetable*}

\section{Best practices for measuring the stellar masses of low-mass galaxies}
\label{sec:bestpractices}

Based on our investigation, we now outline recommendations for measuring the stellar masses of low-mass galaxies ($5<\log(M_{\star}/\rm{M_{\odot}})<8$).

\begin{enumerate}
    \item If multi-wavelength SED fitting is not computationally or observationally feasible, we advise caution when using literature color--$M_\star/L$ relations, which are typically derived using templates of smooth SFHs. 
    For low-mass galaxies with bursty SFHs, we propose a new $g-r$-based calibration (Equation~\ref{eq:ourcalib_opt}) to calculate the stellar masses of dwarf galaxies ($M_{g,0}>-18.5$) from optical fluxes.
    Based on our simulated sample, this color--$M_\star/L$ relation is less systematically biased as a function of mass than literature relations, and can recover stellar masses with $1\sigma$ errors of $\sim0.1$~dex (Figure~\ref{fig:newphotcalib}).
    This is comparable to the best results from multi-wavelength SED fitting, and may significantly outperform SED fitting with certain choices of assumptions (see below). However, we note that Equation~\ref{eq:ourcalib_opt} is (like all empirical calibrations) only as good as the data used to calibrate it!
    We cannot guarantee that the simulated galaxies used in this work are representative of real low-mass galaxies.
    \item If possible, we recommend using multi-wavelength SED fitting assuming a \emph{non-parametric} SFH (Figure~\ref{fig:sim_sfhs}). In a best-case scenario, SED fitting with a non-parametric SFH can produce stellar mass estimates with $1\sigma$-level dispersions of $\sim0.1$~dex. 
    \item For SED fitting with a non-parametric SFH, we recommend using at least six logarithmically spaced age bins, and applying some restrictions to prevent extremely large variations between SFH bins (i.e., using a Dirichlet SFH template with $\alpha\geq0.7$ or a Continuity SFH template).
    \item Allowing dust attenuation parameters to vary in an SED fit further increases the dispersion in $M_{\star,\mathrm{calib}}-M_{\star,\mathrm{true}}$ (Figure~\ref{fig:sim_dustatt}) and may potentially introduce uncertainties in individual stellar mass estimates of up to $\sim0.6$~dex (Figure~\ref{fig:dustcheck}), depending on how badly the total dust content is over- or under-estimated. For low-mass galaxies that are not likely to be very dusty, we therefore recommend using narrow prior distributions (e.g., a \citet{Calzetti2000} dust attenuation law with normalization $0<\tau_V<0.4$) to avoid significantly overestimating the amount of dust.
    \item Some SED fitting assumptions do not appear to strongly impact stellar mass recovery. For example, dust emission parameters do not significantly affect $M_\star$ measurements based on optical and near-IR photometry. Changing the assumed stellar IMF may introduce constant offsets to $M_\star$ measurements, but these are relatively small ($<0.2$~dex) for low-mass galaxies.
    \item If attempting to choose a limited subset of photometric bands to observe, of the bands we tested (NUV, $grz$, and WISE W1/W2) the optical $grz$ bands provide the most information about $M_\star$ for low-mass galaxies. While adding more bands can further improve stellar mass estimates, any bands with large observational uncertainties (e.g., WISE W2 in this study) may actually increase the uncertainty in $M_\star$. 
\end{enumerate}

\section{Discussion}
\label{sec:discussion}

\subsection{Differences between models}
\label{sec:sim_differences}
A fundamental flaw of the tests presented in Sections~\ref{sec:results_photcalib} and \ref{sec:results_sed} is our inherent assumption that simulated galaxies accurately represent real low-mass galaxies. 
We have attempted to mitigate this by using several different galaxy models that use a variety of approaches and physical assumptions, as described in Section~\ref{sec:models}.
Here we discuss whether our results depend strongly on these models.

For many of the empirical photometric calibrations discussed in Section~\ref{sec:results_photcalib} and most of the SED fitting tests discussed in Section~\ref{sec:results_sed}, the mean residuals computed for the low-mass galaxies from each of the four models (GRUMPY, UM-SAGA, MARVEL, FIRE) typically agree within 1$\sigma$ (as seen in each row of Tables~\ref{tab:photresiduals} and \ref{tab:residuals}).
The only exceptions are the UM-SAGA sample, which sometimes has systematically higher stellar mass residuals than the other models.
For example, when the \citet{Klein2024} empirical calibration is used to compute stellar masses, the UM-SAGA sample has an average stellar mass residual of $0.33\pm0.03$~dex, compared to $0.18\pm0.08$~dex for the GRUMPY sample and $0.16\pm0.11$~dex for the FIRE sample.
Similarly, when testing SED fits with a Dirichlet SFH model, the UM-SAGA galaxies have a average stellar mass residual of $0.04\pm0.04$, compared to $-0.08\pm0.05$ for the GRUMPY sample.

In the case of SED fitting, the discrepancy in the mass calibration is likely the result of the small sample ($N=7$) of UM-SAGA galaxies used in the analysis. When we include low-mass ($<10^{7}~\rm{M_{\odot}}$) UM-SAGA galaxies in the analysis, which increases the sample size to 42 galaxies\footnote{Of the original $N=43$ UM-SAGA galaxies, one has a stellar mass $>10^{8}~\rm{M_{\odot}}$ and is excluded from the mean and standard deviation residuals listed in Tables~\ref{tab:photresiduals} and \ref{tab:residuals}.}, nearly all of the discrepancies in mass residuals from SED fitting disappear. 
When empirical photometric calibrations are used, on the other hand, the mass residual discrepancies persist even when the larger UM-SAGA galaxy sample is used.
It is likely that the empirical photometric calibrations struggle to capture the late-time bursty behavior in the SFHs, which exists in many UM-SAGA galaxies (Figures~\ref{fig:sfh_sim} and \ref{fig:samplesfh}).

In short, the discrepancies between the different galaxy models used in this work further indicate that the performance of any stellar mass measurement technique is strongly dependent on how well its assumptions about SFHs match actual galaxy SFHs.
However, we point out that the discrepancies between the stellar mass residuals $\log(M_{\star,\mathrm{calib}}/M_{\star,\mathrm{true}})$ for different simulations are relatively small ($\lesssim0.1$~dex), confirming that stellar mass estimates are generally robust.

\subsection{Additional systematics}
\label{sec:systematics}
We now discuss systematic effects that may impact the applicability of our results.
Some of these involve physics that have been simplified or ignored in modeling low-mass galaxies, while other systematic effects are due to our experimental setup (i.e., fixing certain variables, choice of SED fitting code).

\subsubsection{Physical factors}
For simplicity, we have ignored a number of physical factors in our modeling of low-mass galaxies.

First, we have overlooked the potential contribution of active galactic nuclei (AGN) to low-mass galaxies \citep[see, e.g., the recent review of][]{Reines2022}.
None of the model galaxies used in this work include AGN, and we fixed the AGN contribution to zero when performing SED fits using Prospector.
In real galaxy surveys, the AGN contribution may not be exactly zero.
This can lead to systematic offsets in $M_\star$ measurements, since ignoring the contribution of an AGN to a galaxy's light can lead to an overestimate of the stellar mass.
The magnitude of this effect is unclear and highly dependent on how AGN are modeled: some studies find that AGN in the local universe do not strongly impact $M_\star$ measurements from SED fitting \citep[e.g.,][]{Leja2018,Thorne2022}, while others suggest that some AGN models (particularly for high-luminosity AGN) may lead to $M_\star$ uncertainties of up to $\sim0.5$~dex \citep[e.g.,][]{Buchner2024,Ciesla2015}.

Most of this literature has focused on higher-mass ($M_\star>10^8~\rm{M_\odot}$) galaxies, so it is even less clear how AGN contributions might affect $M_\star$ measurements in the low-mass regime.
Estimates of the fraction of low-mass galaxies in the nearby universe with AGN vary wildly---while some studies suggest that the AGN occupation fraction is relatively small \citep[$\lesssim0.5\%$ in galaxies with $M_\star\lesssim10^9$ out to redshifts $z<0.1$; e.g.,][]{Birchall2020,Mezcua2018}, others find AGN fractions of up to 2\% \citep{Pucha2025}, $\gtrsim20\%$ \citep{Mezcua2024}, or higher \citep{Burke2025}.
It is difficult to overstate just how little we know about the effects of AGN on low-mass galaxies, and additional modeling of AGN in low-mass galaxies is crucial to better constrain uncertainties on measurements of physical properties.

Another physical phenomenon that may impact low-mass galaxies is quenching from reionization.
We briefly consider how different assumptions about reionization might affect the stellar masses of the mock galaxies in our sample.
The galaxy models described in Section~\ref{sec:models} parameterize reionization in a variety of ways.
In the semi-analytic model GRUMPY, UV heating from reionization is parameterized as a suppression factor on gas inflow, while in the zoom-in simulations MARVEL and FIRE, reionization is directly implemented as a photoionizing UV background \citep[described respectively in][]{Haardt12,Faucher-Giguere2009}.\footnote{UM-SAGA does not model reionization at all, since its primary sample consists of galaxies with $M_\star>10^7~\rm{M_\odot}$ that are massive enough to be less impacted by reionization.}
While a full discussion of all the differences among these models is beyond the scope of this paper, we focus on one major difference in the \emph{timing} of reionization in each model.
The redshift of reionization is set to $z_\mathrm{reioniz}=6$ in GRUMPY, $z_\mathrm{reioniz}\sim15$ in MARVEL, and $z_\mathrm{reioniz}\sim10$ in FIRE.
This may have a direct effect on stellar mass: to first order, reionization occurring later leaves more time to form stars before quenching, leading to higher $M_\star$.

To check this, we would ideally look at the galaxies in our sample with $M_\star\lesssim10^6~\rm{M_\odot}$, since these are thought to be the most strongly impacted by reionization \citep[e.g.,][]{Wheeler2019}.
However, our sample in this mass regime is limited --- both GRUMPY and MARVEL suffer from small sample sizes in the $10^{4-6}~\rm{M_\odot}$ mass range ($N=5$ and $N=1$, respectively, as shown in the top panel of Figure~\ref{fig:sfh_sim}), and only a few GRUMPY galaxies have stellar masses $<10^5~\rm{M_\odot}$ (Figure~\ref{fig:gr_sim}) --- making it difficult to identify any differences among simulations.
A larger sample of model galaxies is needed to fully test whether stellar mass recovery depends on $z_\mathrm{reioniz}$. 
Additionally, we note that all models in this work assume uniform reionization for all galaxies.
A potentially more realistic model of ``patchy'' reionization \citep[e.g.,][]{Pentericci2014} would imply that low-mass galaxies in denser environments should undergo reionization quenching first; in other words, $z_\mathrm{reioniz}$ may also be a function of a galaxy's environment.
As described in Section~\ref{sec:models}, nearly all of the model galaxies in our sample are satellite galaxies.
Testing the effect of uniform vs. patchy reionization would require not only a larger sample of $<10^6~\rm{M_\odot}$ galaxies, but also a sample that spans a wide range of environments (satellite, field, and isolated).

\subsubsection{Experimental setup}
\label{sec:expsetup}

Other systematic effects may be due not to physical factors, but our choices when setting up our tests of stellar mass measurement techniques.

For example, throughout this work, we assumed the redshifts of our mock galaxies were exactly known.
While this is a reasonable assumption for galaxy surveys that obtain precise spectroscopic redshifts, it may have a larger impact on stellar mass measurements obtained from purely photometric surveys of low-mass galaxies.
Over- or underestimating a galaxy's redshift will produce systematic offsets in stellar mass.
The magnitude of this effect can be estimated with simple scaling relations: for $z\ll1$, $z\propto d$ and luminosity $L\propto Fd^2$ (where $d$ is luminosity distance and $F$ is observed flux).
Assuming $M_\star\propto L$ (which is generally reasonable, as shown by the empirical calibrations in Section~\ref{sec:calibrations}), this leads to $\log(M_{\star,\mathrm{calib}}/M_{\star,\mathrm{true}})\approx2\log(z_{\mathrm{calib}}/z_{\mathrm{true}})$.
For a galaxy with $z_\mathrm{true}\approx0.1$, typical photometric redshift errors of $\sim0.03$ \citep{Salvato2019} can produce stellar mass residuals of $\sim0.3$~dex.
This is an appreciable uncertainty, particularly when compared to the $1\sigma$ dispersions of $\sim0.1$~dex that we found for our optical/NIR photometric calibrations and SED fitting (Section~\ref{sec:bestpractices}).\footnote{We emphasize that this estimate is only valid when $z\ll1$; as redshift increases, the choice of cosmology also begins to contribute to the uncertainty in stellar mass.}

Furthermore, for most of the SED fitting tests in Section~\ref{sec:prospector}, redshift was not the only parameter we held fixed.
We treated the SFH prior, dust attenuation, dust extinction, and IMF as independent assumptions. 
By only varying the assumption under investigation and fixing all other parameters, we may have ignored covariant effects. 
The true uncertainties in stellar mass likely depend on how assumptions about the SFH, dust, and IMF are combined.

Finally, as discussed in Section~\ref{sec:prospector}, we chose Prospector as our SED fitting code because it uses FSPS to create galaxy models.
Since our mock observations were produced using FSPS, our experiment is therefore a direct ``apples-to-apples'' comparison that tests, to first order, the SED fitting mechanism rather than other systematic effects that may depend on the exact choice of SED fitting code (e.g., stellar model libraries, dust parameterization, SED fitting algorithm).
However, when fitting the SEDs of real galaxies, there is no way to select a code that exactly matches the input SED, so it is important to understand the systematic uncertainties introduced by different SED fitting codes.
\citet{Pacifici2023} recently investigated the effects of using 14 different SED fitting codes on high-mass ($>10^8~M_\odot$) galaxies at $z\sim1$ and $z\sim3$ and estimate that the choice of code contributes $\sim0.12$~dex to the uncertainty in stellar mass.
We plan to extend this test to the low-mass regime in a future study.

\section{Summary}
\label{sec:summary}

As we enter an era of unprecedentedly large extragalactic surveys, we are beginning to observe statistical samples of low-mass galaxies across a range of environments. 
Accurately estimating the physical properties of these galaxies is critical if we want to understand the physics that drives their evolution.
However, many of the methods used to measure galaxy properties from integrated light have primarily been developed for and tested on massive ($>10^8~\rm{M_\odot}$) galaxies.
In this study, we have attempted to verify several of these methods for low-mass galaxies.

In particular, we have tested how well different stellar mass measurement techniques can recover the masses of low-mass galaxies from integrated UV/optical/near-IR photometry.
Our test sample included 469 simulated galaxies from four different models.
Using multiple types of simulations --- semi-analytic models, empirical halo-galaxy connection models, and two different hydrodynamic simulations with varying resolutions and prescriptions for feedback --- makes our results more robust to variations among different simulations.
For each simulated galaxy, we generated mock photometry using the stellar population synthesis code FSPS.

We then tested empirical color--$M_\star/L$ calibrations and showed that many literature relations produce systematic trends in stellar mass residuals that are only apparent at the low-mass end.
This is particularly true for calibrations based only on near-IR color, which can produce stellar mass errors of $>1$~dex for galaxies below $10^{8}~\rm{M_{\odot}}$.
We have provided an updated prescription for stellar mass based on $g-r$ color (Equation~\ref{eq:ourcalib_opt}) that reduces these systematic biases.
This new calibration can recover stellar masses with minimal offsets ($0.03\pm0.07$~dex). 

We also tested different assumptions that go into multi-wavelength SED fitting.
In agreement with previous studies of high-mass galaxies \citep{Lower2020}, we found that non-parametric SFH models generally perform better than parametric SFH models when measuring stellar masses of low-mass galaxies.
While parametric SFHs can underestimate stellar mass by as much as $\sim0.4$~dex, under ideal conditions non-parametric SFH templates can recover $M_\star$ of low-mass galaxies with offsets of $-0.03\pm0.09$~dex.
Assumptions about dust attenuation parameters introduce larger uncertainties in $M_\star$ of low-mass galaxies: over- or under-estimating total dust content can lead to significant ($\sim0.6$~dex) over- or under-estimates of stellar mass.
The shape of the dust attenuation law may also impact stellar mass estimates, though more work is needed to better understand these effects; in this study we only compared two simple attenuation models with different UV-optical slopes, and we did not consider other features in the attenuation curve \citep[e.g., the UV bump;][]{Salim2020}. 

Our results are summarized in Section~\ref{sec:bestpractices}, in which we have laid out recommendations for measuring $M_\star$ of low-mass galaxies from integrated photometry.
In general, we recommend using multi-wavelength SED fitting with the assumption of a non-parametric SFH.

This study represents an initial step towards a full re-evaluation of measurement techniques based on the integrated light of low-mass galaxies.
Much remains to be done, including further tests of SED fitting.
As discussed in Section~\ref{sec:systematics}, this work aimed to test the assumptions that go into a single SED fitting code (Prospector, which uses the same SPS models that were used to generate our mock observations).
This work also focused on stellar mass as a ``zeroth-order'' property that traces a galaxy's integrated history.
In the future, we plan to compare other SED and spectral fitting codes, as well as to investigate other galaxy properties --- star formation rates and histories, metallicities, AGN properties --- that can reveal even more information about the physics underlying low-mass galaxy evolution.

\acknowledgments
{MAdlR acknowledges the financial support of the Stanford Science Fellowship while writing this paper. 
This research has made use of NASA’s Astrophysics Data System Bibliographic Services.

There are many communities without whom this work would not have been possible. 
We acknowledge that this work is rooted in Western scientific practices and is the material product of a long and complex history of settler-colonialism. MAdlR wishes to recognize her status as a settler on the unceded homelands of the Pocumtuc Nation.
We hope to work toward a scientific practice guided by pono and a future in which we all honor the land.

This research used data from the SAGA Survey (Satellites Around Galactic Analogs; sagasurvey.org). The SAGA Survey is a galaxy redshift survey with spectroscopic data obtained by the SAGA Survey team with the Anglo-Australian Telescope, MMT Observatory, Palomar Observatory, W. M. Keck Observatory, and the South African Astronomical Observatory (SAAO). The SAGA Survey also made use of many public data sets, including: imaging data from the Sloan Digital Sky Survey (SDSS), the Dark Energy Survey (DES), the GALEX Survey, and the Dark Energy Spectroscopic Instrument (DESI) Legacy Imaging Surveys, which includes the Dark Energy Camera Legacy Survey (DECaLS), the Beijing-Arizona Sky Survey (BASS), and the Mayall z-band Legacy Survey (MzLS); redshift catalogs from SDSS, DESI, the Galaxy And Mass Assembly (GAMA) Survey, the Prism Multi-object Survey (PRIMUS), the VIMOS Public Extragalactic Redshift Survey (VIPERS), the WiggleZ Dark Energy Survey (WiggleZ), the 2dF Galaxy Redshift Survey (2dFGRS), the HectoMAP Redshift Survey, the HETDEX Source Catalog, the 6dF Galaxy Survey (6dFGS), the Hectospec Cluster Survey (HeCS), the Australian Dark Energy Survey (OzDES), the 2-degree Field Lensing Survey (2dFLenS), and the Las Campanas Redshift Survey (LCRS); HI data from the Arecibo Legacy Fast ALFA Survey (ALFALFA), the FAST all sky HI Survey (FASHI), and HI Parkes All-Sky Survey (HIPASS); and compiled data from the NASA-Sloan Atlas (NSA), the Siena Galaxy Atlas (SGA), the HyperLeda database, and the Extragalactic Distance Database (EDD). The SAGA Survey was supported in part by NSF collaborative grants AST-1517148 and AST-1517422 and Heising–Simons Foundation grant 2019-1402. SAGA Survey's full acknowledgments can be found at \url{https://sagasurvey.org/ack/}. }

\vspace{5mm}

\software{
Prospector \citep{Johnson2021,Leja2019},
python-FSPS \citep{pyfsps},
FSPS \citep{Conroy2009,Conroy2010},
Matplotlib \citep{matplotlib}, 
Seaborn \citep{seaborn}, 
Astropy \citep{astropy},
Scipy \citep{scipy}}

\appendix

\section{Prospector priors}
\label{sec:priors}

For the sake of reproducibility, the full priors for all Prospector runs used in this study are described in Table~\ref{tab:priors}.

\begin{deluxetable*}{llll}[h!]
\tablecolumns{4} 
\tablecaption{Prior distributions used to fit SED models. \label{tab:priors}} 
\tablehead{ 
\multicolumn{4}{c}{For all tests} \\
\colhead{} & \colhead{Model} & \colhead{Parameter} & \colhead{Prior}
}
\startdata
 All models & $M_{\star}-Z_{\star}$ & Stellar mass formed $\log(M_{\star}/M_{\odot})$ & Uniform(4,11)\\
 & & Stellar metallicity $\log(Z_{\star}/Z_{\odot})$ & Uniform(-3.5, 0.19) \\
\hline
\hline
\multicolumn{4}{c}{Parametric vs. non-parametric SFH test} \\
\colhead{} & \colhead{Model} & \colhead{Parameter} & \colhead{Prior} \\
\hline
 All models & \citet{Chabrier2003} IMF & & Fixed \\
 & \citet{Draine2007} dust emission & Minimum radiation field & Fixed at 1.0 \\
 & & Warm dust fraction $\gamma$ & Fixed at 0.01 \\
 & & PAH mass fraction $q_{\mathrm{PAH}}$ & Fixed at 3.5\% \\
 & \citet{Calzetti2000} attenuation & Optical depth at 5500\AA\ $\tau_{V}$ & Fixed at 0.2 \\
Parametric SFHs & Constant & n/a & n/a \\
 & Delayed-$\tau$ & $T_0$ & Uniform(0.01, 13.8) Gyr \\
 & & $\tau$ & LogUniform(0.001, 10) Gyr$^{-1}$ \\
 & Delayed-$\tau$ + burst & $T_0$, $\tau$ & as above \\
 & & Burst time & Uniform(0.5, 1.0) * Age \\
 & & Burst $M_{\star}$ fraction & Uniform(0.0, 5.0) * $M_{\star}$ \\
Non-parametric SFHs & Log$M$ & $\log(M_{\star}/M_{\odot})$ per bin & Uniform(4,11) \\
& Continuity & log(SFR) ratios & Student's t: loc=0, scale=0.3, $\nu$=2\\
& Dirichlet &
sSFR for each time bin & Dirichlet: $\alpha=0.7$ \\
\hline
\hline
\multicolumn{4}{c}{Dust attenuation test} \\
\colhead{} & \colhead{Model} & \colhead{Parameter} & \colhead{Prior distributions} \\
\hline
 All models & \citet{Chabrier2003} IMF & & Fixed \\
 & Dirichlet SFH & sSFR for each time bin & Dirichlet: $\alpha=0.7$ \\
 & \citet{Draine2007} dust emission & Minimum radiation field & Fixed at 1.0 \\
 & & Warm dust fraction $\gamma$ & Fixed at 0.01 \\
 & & PAH mass fraction $q_{\mathrm{PAH}}$ & Fixed at 3.5\% \\
Calzetti free & \citet{Calzetti2000} attenuation & Optical depth at 5500\AA\ $\tau_{V}$ & Uniform(0,2) \\
SMC & \citet{Gordon2003} extinction & Young stellar attenuation $\tau_{\mathrm{young}}$ & Fixed at 1.0 \\
 & & Diffuse attenuation $\tau_{\mathrm{diff}}$ & Fixed at 0.3 \\
\hline
\hline
\multicolumn{4}{c}{Dust emission test} \\
\colhead{} & \colhead{Model} & \colhead{Parameter} & \colhead{Prior distributions} \\
\hline
 All models & Dirichlet SFH & sSFR for each time bin & Dirichlet: $\alpha=0.7$ \\
 & \citet{Calzetti2000} attenuation & Optical depth at 5500\AA\ $\tau_{V}$ & Fixed at 0.2 \\
 & \citet{Draine2007} dust emission & Minimum radiation field & Uniform(0.1,25.0) \\
 & & Warm dust fraction $\gamma$ & Uniform(0.0,1.0) \\
 & & PAH mass fraction $q_{\mathrm{PAH}}$ (\%) & Uniform(0.0,10.0) \\
\hline
\hline
\multicolumn{4}{c}{IMF test} \\
\colhead{} & \colhead{Model} & \colhead{Parameter} & \colhead{Prior distributions} \\
\hline
 All models & Dirichlet SFH & sSFR for each time bin & Dirichlet: $\alpha=0.7$ \\
 & \citet{Calzetti2000} attenuation & Optical depth at 5500\AA\ $\tau_{V}$ & Fixed at 0.2 \\
 & \citet{Draine2007} dust emission & Minimum radiation field & Fixed at 1.0 \\
 & & Warm dust fraction $\gamma$ & Fixed at 0.01 \\
 & & PAH mass fraction $q_{\mathrm{PAH}}$ & Fixed at 3.5\% \\
 Kroupa & \citet{Kroupa2001} IMF & & Fixed \\
 Salpeter & \citet{Salpeter1955} IMF & & Fixed \\
\enddata
\end{deluxetable*}

\bibliographystyle{aasjournal}
\bibliography{main}

\end{document}